\documentclass[12pt,journal,draftcls,a4paper,onecolumn]{IEEEtran}

\usepackage[nomarkers,nofiglist,notablist]{endfloat} 

\usepackage{dsfont}
\usepackage{amsmath}
\usepackage{amssymb}
\usepackage{amsfonts}
\usepackage{graphicx}
\usepackage{amsmath}
\usepackage[all,poly]{xy}
\usepackage{multirow}
\usepackage{algorithm}
\usepackage{algorithmic}
\usepackage{color}
\usepackage[noadjust]{cite}
\usepackage{subfigure}
\usepackage{tikz,pgf}
\usetikzlibrary{arrows,automata}
\usepackage{verbatim}
\usetikzlibrary{positioning,shapes.geometric}
\usepackage{url}
\usepackage[T1]{fontenc}

\newcommand{\figwidth}{\columnwidth}

\newcommand{\bone}{\boldsymbol{1}}
\newcommand{\bzero}{\boldsymbol{0}}

\newcommand{\bSig}{\boldsymbol{\Sigma}}
\newcommand{\bgam}{\boldsymbol{\gamma}}
\newcommand{\bGam}{\boldsymbol{\Gamma}}

\newcommand{\bphi}{\boldsymbol{\phi}}
\newcommand{\bPhi}{\boldsymbol{\Phi}}
\newcommand{\red}{\textcolor[rgb]{1.00,0.00,0.00}}
\newcommand{\blue}{\textcolor[rgb]{0.00,0.00,0.80}}

\def\bsa{{\boldsymbol{a}}}
\def\bsb{{\boldsymbol{b}}}

\def\bse{{\boldsymbol{e}}}

\def\bsk{{\boldsymbol{k}}}

\def\bsm{{\boldsymbol{m}}}

\def\bsp{{\boldsymbol{p}}}
\def\bsq{{\boldsymbol{q}}}

\def\bss{{\boldsymbol{s}}}

\def\bsu{{\boldsymbol{u}}}
\def\bsv{{\boldsymbol{v}}}

\def\bsx{{\boldsymbol{x}}}
\def\bsy{{\boldsymbol{y}}}

\def\bsA{{\boldsymbol{A}}}
\def\bsB{{\boldsymbol{B}}}

\def\bsD{{\boldsymbol{D}}}

\def\bsF{{\boldsymbol{F}}}
\def\bsG{{\boldsymbol{G}}}
\def\bsH{{\boldsymbol{H}}}

\def\bsM{{\boldsymbol{M}}}

\def\bsP{{\boldsymbol{P}}}
\def\bsQ{{\boldsymbol{Q}}}

\def\bsU{{\boldsymbol{U}}}
\def\bsV{{\boldsymbol{V}}}

\def\bsX{{\boldsymbol{X}}}
\def\bsY{{\boldsymbol{Y}}}
\def\bsZ{{\boldsymbol{Z}}}

\def\dsR{{\mathds{R}}}

\def\calN{{\mathcal{N}}}

\newcommand{\diag}[1]{\mathrm{diag}{\left\{#1\right\}}}

\newcounter{algo}
\renewcommand{\thealgo}{\arabic{algo}}

\title{Fast Hyperspectral Unmixing in Presence of Nonlinearity or Mismodelling Effects}

\author{Abderrahim Halimi$^{\,1}$\thanks{(1) A. Halimi, G. S. Buller and S. McLaughlin  are with the School of Engineering and Physical Sciences, Heriot-Watt University, Edinburgh U.K. (e-mail: \{a.halimi, g.s.buller, s.mclaughlin\}@hw.ac.uk).}, Jose Bioucas-Dias$^{\,2}$\thanks{(2) J. M. Bioucas-Dias is with  the  Instituto  de  Telecomunica\c{c}\~oes  and  Instituto  Superior  T\'ecnico,  Universidade  de  Lisboa, Portugal  (e-mail: bioucas@lx.it.pt)}, Nicolas Dobigeon$^{\,3}$\thanks{(3) N. Dobigeon is with the University of Toulouse, 2 rue Charles Camichel, BP 7122, 31071 Toulouse cedex 7,
France (e-mail: Nicolas.Dobigeon@enseeiht.fr).},  Gerald S. Buller$^{\,1}$, Steve McLaughlin$^{\,1}$, 
\thanks{This work was supported by the EPSRC Grants EP/J015180/1, EP/N003446/1, and EP/K015338/1}}

\begin{document}

\maketitle

\begin{abstract}
This paper presents two novel hyperspectral mixture models and associated unmixing algorithms. The two models assume a linear mixing model corrupted by an additive term whose expression can be adapted to account for multiple scattering nonlinearities (NL), or mismodelling effects (ME). The NL model generalizes bilinear models by taking into account higher order interaction terms. The ME model accounts for different effects such as endmember variability or the presence of outliers. The abundance and residual parameters of these models are estimated by considering a convex formulation suitable for fast estimation algorithms. This formulation  accounts for  constraints such as the  sum-to-one and non-negativity of the abundances, the non-negativity of the nonlinearity coefficients, the spectral smoothness of the ME terms and the spatial sparseness of the residuals. The resulting convex problem is solved using the alternating direction method of multipliers (ADMM) whose convergence is ensured theoretically. The proposed mixture models and their unmixing algorithms are validated on both synthetic and real images showing competitive results regarding the quality of the inference and the computational complexity when compared to the state-of-the-art algorithms.
\end{abstract}

\begin{keywords}
Hyperspectral imagery, collaborative sparse regression, ADMM, nonlinear unmixing, robust unmixing, convex optimization
\end{keywords}

\section{Introduction} \label{sec:Introduction} 
Hyperspectral imaging is a remote sensing technology that collects three dimensional data cubes composed of $2$D spatial images acquired in numerous contiguous spectral bands. Assuming that each pixel spectrum is a mixture of several pure materials (endmembers), spectral unmixing consists of recovering the spectral signatures (endmembers) of the materials present in the scene, and quantifying their proportions within each hyperspectral image pixel \cite{Bioucas2012}. More precisely, unmixing hyperspectral images consists of three stages: (i) determining the number of endmembers and possibly projecting the data onto a subspace of reduced dimension \cite{HalimiTGRS2016,Bioucas2008,ChangTGRS2004}, (ii) identifying the endmembers using an endmember extraction algorithm (EEA) such as vertex component analysis (VCA) \cite{Nascimento2005},  and N-FINDR \cite{Winter1999}  and (iii) estimating their abundances \cite{Heinz2001,Chen2014,BioucasWhispers2010}. Akin to \cite{Heinz2001,BioucasWhispers2010,AltmannTCI2015b,ChenTSP2013}, this paper considers a supervised unmixing scenario which aims at estimating the abundances while assuming that the two first unmixing steps have been successfully implemented.   

As a result of its simplicity, the linear mixing model (LMM) is used by many of the hyperspectral unmixing algorithms presented in the literature \cite{Heinz2001,BioucasWhispers2010}.  
This is generally justified when considering flat scenes without component interactions, and a fixed endmember spectra for all the pixels. However, an inherent limitation of the LMM occurs in presence of volumetric scattering, terrain relief, or intimate mixtures of materials which require the definition of new sophisticated models, to take  these effects into account. Nonlinear mixture models are an alternative to better account for those effects \cite{Heylen2014,Dobigeon2014} and we distinguish between two main families: the first is signal processing based and seeks to construct flexible models that can represent a wide range of nonlinearities. The second is physical based models that include the intimate mixture models \cite{Hapke1981} and those accounting for bilinear interactions   \cite{Altmann2012,Halimi2011TGRS,HalimiIGARSS2011,Nascimento2009,Fan2009,MeganemTGRS2014}. This paper considers a physical based nonlinearity which generalizes the bilinear formulation in  \cite{HalimiTIP2016,AltmannTCI2015b} to account for multiple scattering effects. A second inherent limitation of the LMM appears when the endmember spectra vary spectrally and spatially causing what is known as endmember variability \cite{Somers2011,Zare2014}. In this case, and under a supervised SU scenario, the endmember fluctuation can not be captured by traditional EEA algorithms which affect the LMM by the presence of an additional spectrally smooth residual component \cite{HalimiTIP2016}. A third LMM limitation is related to the presence of sparse outliers, e.g. due to the presence of impulse noise, horizontal or vertical line stripes, dead lines, and others types of noise \cite{AggarwalJSTARS2016,HeJSTARS2016}. The latter two LMM limitations can be solved separately by considering specialized algorithms that deal with EV \cite{Eches2010ip,Zare2013,Somers2012}   or  outliers \cite{AggarwalJSTARS2016,HeJSTARS2016}. In this paper, we adopt the same strategy as in \cite{HalimiTIP2016,AltmannTCI2015} and propose a robust algorithm that encompasses the first two effects described above.

The first contribution of this paper is the introduction of two mixture models to deal with NL and ME.  The models proposed are based on the residual component principle \cite{KalaitzisICML2012} and are closely relate to the  RCA-NL and RCA-ME models introduced in \cite{HalimiTIP2016}.  More precisely, the proposed NL generalizes RCA-NL by accounting for multiple scattering effects. Indeed, the residual term is assumed to be a linear combination of high order interaction spectra.  Due to the high number of interactions, the non-negative nonlinearity coefficients are assumed sparse so that only a few interactions are active for each pixel. The resulting formulation is then general, and covers many NL  models \cite{Altmann2012,Halimi2011TGRS,HalimiIGARSS2011,Nascimento2009,Fan2009,MeganemTGRS2014,Altmann2014}. 
In a similar fashion to RCA-ME, the proposed ME model assumes a spectrally smooth residual term. However, in contrast with RCA-ME that adopts a statistical approach to account for this smooth property, the proposed model assumes the residual term to be a sparse linear combination over a dictionary (such as the discrete cosine transform (DCT), or the spline decomposition). For both models, the corrupted pixels are assumed spatially sparse meaning that only a small number of nonlinear or outlier pixels are present, as previously suggested in \cite{FevotteTIP2015,Altmann2014} for NL and in  \cite{AggarwalJSTARS2016,HeJSTARS2016} for outliers. This effect has been introduced by considering the well known collaborative sparse regression strategy \cite{SprechmannIEEETSP2011,FevotteTIP2015,AggarwalJSTARS2016,IordacheTGRS2014,IordacheTGRS2014b}  since it  promotes group-sparsity over the residual terms while using the information of the residuals in all the pixels. Note that the first motivation for these new reformulations is that both models assume a residual term that is written as a linear combination of sparse coefficients, which is suitable for the development of a joint formulation to achieve the unmixing strategy. The second motivation is related to the unmixing problem that is significantly simplified by considering separable variables (between the abundances and the residual coefficients) as well as a linear expression for both the LMM term and the residual term.

The second contribution of this paper is the introduction of a convex formulation for unmixing the proposed observation models. The convexity is obtained thanks to the linearity of the observation models with respect to the unknown parameters, as well as the considered regularization terms. Indeed, the formulation accounts for the known physical constraints on the estimated parameters such as the sum-to-one and non-negativity of the abundances, the non-negativity of the nonlinearity coefficients, the spectral smoothness of the ME terms and the spatial sparseness of the residuals.  
The resulting convex problem is solved using the alternating direction method of multipliers (ADMM) whose convergence is theoretically ensured. More precisely, we propose two algorithms denoted as NUSAL-$K$ for nonlinear unmixing by variable splitting and augmented Lagrangian with order $K$, and RUSAL for robust unmixing by variable splitting and augmented Lagrangian. Note that the ADMM  algorithms are well adapted for large scale problems, i.e., with a large number of parameters to be estimated \cite{Afonso_TIP2011,Boyd_FTML2011}. Moreover, this method offers good performance at a reduced computational cost as already shown in many hyperspectral unmixing works \cite{BioucasWhispers2010,IordacheTGRS2014,IordacheTGRS2014b}.
The proposed mixture models and estimation algorithms are validated using synthetic and real
hyperspectral images. The results obtained are very promising and show the potential of the proposed mixture models and associated inference algorithms with respect to the estimation quality and the computational cost.

The paper is structured as follows. Section \ref{sec:Mixture_models} presents the proposed NL and  ME mixture models considered in this study. Section \ref{sec:Proposed_hyperspectral_unmixing_algorithms} introduces the convex unmixing formulations and the ADMM-based optimization algorithms associated with the two mixture models. Section \ref{sec:Simulation_results_on_synthetic_data} analyzes the performance of the proposed algorithms when applied to synthetic images with known ground truth. Results on real hyperspectral images are presented in Section \ref{sec:results_on_real_data} and conclusions and future work are reported in Section
\ref{sec:Conclusions}.
 
\section{Mixture models} \label{sec:Mixture_models}  
As a result of its simplicity, the LMM is widely used in hyperspectral images. However, the LMM has some limitations in the presence of nonlinearity or outlier effects. This paper deals with these issues by considering the observation model proposed in  \cite{HalimiTIP2016}, itself inspired from the residual component analysis model described in \cite{KalaitzisICML2012}. This model introduces a general formulation that is expressed as the sum of a linear model and a residual term that accounts for the remaining effects. The general observation model for the $\left(L \times 1\right)$ pixel spectrum $\bsy_{n}$, where $L$ is the number of spectral bands,  is given by
\begin{eqnarray} 
\bsy_{n} & = &  \sum_{r=1}^{R}{a_{r,n} \bsm_{r} } + \bphi_{n} \left(\bsM,\bsa_{n},\bsx_{n}\right) + \bse_{n}  \nonumber \\
& = &   \bsM  \bsa_{n} +  \bphi_{n}\left(\bsM,\bsa_{n},\bsx_{n}\right) + \bse_{n},
 \label{eqt:GRCA_NL_EV_ME}
\end{eqnarray}
where $\bsa_{n}= \left(a_{1,n},\cdots,a_{R,n} \right)^T$ is an $\left(R\times1\right)$ vector of abundances associated with the $n$th pixel, $\bsx_{n}= \left(x_{1,n},\cdots,x_{D,n} \right)^T$ is a  $\left(D\times1\right)$ vector of residual coefficients associated with the $n$th pixel, $R$ (resp. $D$) is the number of endmembers (resp. residual coefficients),  $\bse_{n}   \sim \calN \left(\bzero ,\bSig  \right)$ is a centered Gaussian noise and $ \bphi_{n}$ is a residual term that might depends on the endmembers, the abundances or  residual coefficients to account for the additional mismodelling effect. In model \eqref{eqt:GRCA_NL_EV_ME}, the endmembers matrix $\bsM$ is fixed (extracted using an EEA) and endmember variability can be accounted for by the pixel dependent residual term $\bphi_{n}$.  Moreover, the paper deals with the supervised case in which we assume the endmembers to be known and we only estimate the abundances and the residual terms. Due to  physical  constraints,  the  abundance  vector $\bsa_{n}$ satisfies the following abundance non-negativity (ANC) and abundance sum-to-one (ASC) constraints
\begin{equation}
a_{r,n} \geq 0, \forall r \in \left\{1,\ldots,R\right\} \quad
\textrm{and} \quad \sum_{r=1}^{R}{a_{r,n}}=1.
\label{eqt:contraints_linear_model}
\end{equation} 
Eq. \eqref{eqt:GRCA_NL_EV_ME} shows a general model that can be adapted to account for different physical phenomena.  The next sections present in details the considered model variants that will account for NL and ME.
  
\subsubsection{Effects of Nonlinearity (NL)} \label{subsec:Nonlinearity_effect}

Nonlinear mixing models provide a powerful tool to deal with the inherent limitations of the LMM. Many nonlinear models have been introduced in the literature and we can divide them into two categories: physical based models (including bilinear and intimate mixture models)  and signal processing models (such as the PPNMM \cite{Altmann2012,AltmannICASSP2011}). This paper considers a physical based model to deal with the multiple scattering effects. More precisely, the model considered accounts for higher order interactions between the endmembers and reduces to \cite{HalimiTIP2016,AltmannTCI2015b} when only the bilinear second order interactions are considered. Note that bilinear models assume that the effect of the interaction terms decreases as the order increases, as suggested in \cite{Halimi2011TGRS,Nascimento2009,Fan2009}. However, in this paper, we include higher order interaction terms in the proposed model/algorithm to highlight their benefit as recently shown in \cite{HeylenTGRS2016}. The proposed NL model considering the $K$th order of interactions is given by
\begin{equation}
\bsy_{n} =  \bsM  \bsa_{n} +  \bphi_{n}^{\textrm{NL-}K}  \left(\bsM,\bgam_{n}\right) + \bse_{n}
 \label{eqt:GRCA_NL}
\end{equation}
where the residual component is  
\begin{equation}
\bphi_{n}^{\textrm{NL-}K} \left(\bsM,\bgam_{n}\right) = \bsQ^{(K)} (\bsM) \bgam_{n},  
 \label{eqt:NL}
\end{equation}
with $\bgam_{n} =\left(\gamma_{n}^{(1)}, \cdots, \gamma_{n}^{(D_K)} \right)^T, \forall n$ 
is the $\left(D_K\times1\right)$ vector of non-negative coefficients (i.e., $\gamma_{n}^{(d)} \geq0, \forall n,d$), $\bsQ^{(K)}$  
is the $\left(L\times D_K\right)$ matrix gathering the interaction spectra of the form  
$\bsm_{i} \odot \bsm_{j} \odot \cdots \odot \bsm_{l}$, $\odot$  denotes the Hadamard (term-wise) product, and
 $D_K$ is the number of coefficients associated with the interaction terms that have an order lower or equal to $K$. More details regarding the construction of $\bsQ^{(K)}$ are provided in Appendix \ref{app:Construction_of_Q}. For instance, considering only second order interaction terms (i.e., $K=2$) leads to $D_2= \frac{R (R+1)}{2}$,  $\bgam_{n}(2)=\left(\gamma_{n}^{(1,2)},\cdots,\gamma_{n}^{(R-1,R)}, \gamma_{n}^{(1,1)}, \cdots, \gamma_{n}^{(R,R)} \right)^T, \forall n$,   $\bsQ^{(2)}(\bsM) = \left(\sqrt{2} \bsm_{12}, \cdots, \sqrt{2} \bsm_{R-1,R}, \bsm_{11}, \cdots, \bsm_{RR}\right),$ and a residual term similar to \cite{AltmannTCI2015b} as follows
\begin{equation}
\bphi_{n}^{\textrm{NL-}2} \left(\bsM,\bgam_{n}\right) = \bsQ^{(2)}(\bsM) \bgam_{n}  = \sum_{r=1}^{R}{\gamma_{n}^{(r,r)}  \bsm_{r,r}   }   +  \sum_{r=1}^{R-1}{\sum_{r'=r+1}^{R}{\gamma_{n}^{(r,r')} \sqrt{2} \bsm_{r,r'} }} 
 \label{eqt:NL2}
\end{equation}  
where  $\bsm_{i,j}=\bsm_{i} \odot \bsm_{j}$,  
and the interaction terms are weighted by the coefficient $\sqrt{2}$   obtained by comparison with a homogeneous polynomial kernel of the $2$nd degree (see Appendix \ref{app:Construction_of_Q} for more details regarding these coefficients).
In what follows, and for brevity, we drop the order index $(K)$ for general statements (related to all interaction orders) and only include it when dealing with specific orders. 
The model proposed in \eqref{eqt:GRCA_NL} reduces to the LMM for $\bgam_{n}=0, \forall n$ and has many links to state-of-the-art models. Indeed, model \eqref{eqt:GRCA_NL} with $K=2$ is similar to \cite{AltmannTCI2015b} and has a close relation to the RCA model \cite{Altmann2014} (as shown in \cite{AltmannTCI2015b}). Moreover, it generalizes the GBM model \cite{Halimi2011TGRS,HalimiIGARSS2011} by accounting for self-interaction between the endmembers, and also generalizes the PPNMM \cite{Altmann2012} by considering different weights for the bilinear terms. Overall, model \eqref{eqt:GRCA_NL} is of a similar polynomial form as the bilinear models (GBM \cite{Halimi2011TGRS}, PPNMM \cite{Altmann2012}, Nascimento \cite{Nascimento2009}, Fan \cite{Fan2009}, and Meganem \cite{MeganemTGRS2014} models) with the main difference due to the introduction of higher order interaction terms, and the non-negativity and sum-to-one constraints associated with each model. In contrast with the model described in \cite{HeylenTGRS2016}, which accounts for all the interactions by using only one parameter, the model \eqref{eqt:GRCA_NL} includes a different coefficient for each interaction term, which enables analysis of the interaction between any specific physical components (i.e., availability of interaction maps).

Note that the nonlinear behavior generally affects some pixels of the image as already exploited in \cite{FevotteTIP2015,Altmann2014}, which suggest a spatial sparsity of the nonlinear pixels. Moreover, it makes sense to assume that the elements of the nonlinear vector $\bgam_{n}$ will not be active at the same time, meaning that the vector is sparse. This can be explained since the lowest order of interactions  have often a higher effect \cite{Halimi2011TGRS,Nascimento2009,Fan2009} and all the interactions between endmembers are not likely to be active at the same time. These sparsity properties are of great importance and will be exploited when designing the unmixing algorithm associated with model \eqref{eqt:GRCA_NL} in Section \ref{sec:Proposed_hyperspectral_unmixing_algorithms}.
 
\subsubsection{Mismodelling effects (ME) or outliers} \label{subsec:Mismodelling_effect_or_outliers}
In recent years, there has been considerable interest in robust hyperspectral unmixing to enable adaptation of the simple LMM to realistic scenes which often present outliers or other unknown effects \cite{HalimiSSPD2016}. This goal can be achieved using different strategies such as adapting the optimization cost function \cite{Zhu_Arxiv2016} or changing the observation model by introducing a residual term  that accounts for the mismodelling effects \cite{HalimiTIP2016,AltmannTCI2015,FevotteTIP2015}. The latter strategy is adopted in this paper by considering spectrally smooth residuals as for the ME model introduced in \cite{HalimiTIP2016}. More precisely, the model is 
\begin{equation}
\bsy_{n} =    \bsM  \bsa_{n} +  \bphi_{n}^{\textrm{ME}}(\bsb_n) + \bse_{n}, 
 \label{eqt:GRCA_RL} 
\end{equation}
where the residual component is
\begin{equation}
\bphi_{n}^{\textrm{ME}}(\bsb_n)  = \bsF^{\top} \bsb_n,
 \label{eqt:GRCA_RL2} 
\end{equation} 
with $\bsF$ is a $D\times L$ matrix gathering the first $D$ rows of the DCT,  $\bsb_n$ is a vector of DCT coefficients and $\bphi_{n}^{\textrm{ME}}$ is a smooth spectral function. In this paper, the smooth property of $\bphi_{n}^{\textrm{ME}}$ is obtained by imposing sparsity on the elements of each vector $\bsb_n, \forall n$. Model \eqref{eqt:GRCA_RL} reduces to the LMM for $\bsb_{n}=\bzero_L,  \forall n$. Moreover, the residual terms  $\left\lbrace \bphi_{1}^{\textrm{ME}}, \cdots,\bphi_{N}^{\textrm{ME}} \right\rbrace$ are assumed to be spatially sparse to approximate sparse nonlinear effects, endmember variability effects or other mismodelling effects such as outliers. In the following, we highlight the link between model \eqref{eqt:GRCA_RL} and each of these phenomena. Consider first the NL model \eqref{eqt:GRCA_NL} with $\gamma_{n}=\gamma_{n}^{(d)}=\gamma_{n}^{(d')}, \forall d\neq d'$. In this special case, the nonlinear term reduces to  $\bphi_{n}^{\textrm{NL}} \left(\bsM\right) =  \gamma_{n} \sum_{d=1}^{D}{\bsq_{d}},$ where $\bsq_{d}$ represents the $d$th column of $\bsQ$. Thus, as a result of the smooth spectral property of the interaction spectra $\bsq_{d}$, the nonlinear term $\bphi_{n}^{\textrm{NL}}$ can be approximated by the term $\bphi_{n}^{\textrm{ME}}$. This means that model \eqref{eqt:GRCA_RL} links to the NL model \eqref{eqt:GRCA_NL} for this special case. Second, the EV-model proposed in \cite{HalimiTIP2016} and assuming pixel dependent endmembers $\bss_{r,n}=\bsm_{r,n}+\bsk_{r,n}$, reduces to model \eqref{eqt:GRCA_RL} when the same variability affects the different endmembers (i.e., $\bsk_{r,n}=\bsk_{r',n}, \forall r\neq r'$). Third, spatially sparse outliers can be present in hyperspectral images as shown in \cite{FevotteTIP2015,HeJSTARS2016,AggarwalJSTARS2016}, and can also be approximated using  $\bphi_{n}^{\textrm{ME}}$. 
This illustrates how the model described by \eqref{eqt:GRCA_RL} can be used to process hyperspectral images with a combination of different effects such as NL, EV and/or outliers.  The next section introduces the proposed estimation algorithms associated with these NL and ME models.

\section{Proposed unmixing algorithms: \\  NUSAL-$K$, and RUSAL} \label{sec:Proposed_hyperspectral_unmixing_algorithms}
\begin{figure*} 
\centering
\includegraphics[width=1.\figwidth]{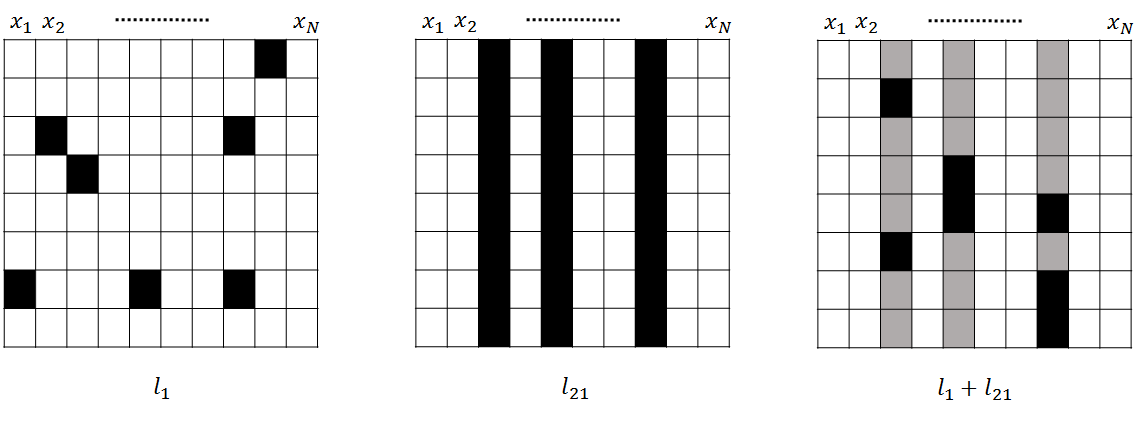}
\caption{Graphical illustration of the performance of the regularization terms. Active members of the matrix $\bsX$ are represented in black, and non-active members are shown in white. The active blocks when considering the $\ell_1+\ell_{21}$ norms are shown in gray.  (left) $\ell_1$ norm, (middle) $\ell_{21}$ norm, and (right) the considered $\ell_1+\ell_{21}$ norms. } \label{fig:Norms_Graph}
\end{figure*}

This section introduces the unmixing algorithms used to estimate the abundances and the residual coefficients of the proposed models. To this end, we adopt an optimization approach that minimizes a regularized data fidelity cost function. More precisely, considering an  independent and identically distributed (i.i.d.)  Gaussian noise ($\bSig$ proportional to the identity matrix) in model \eqref{eqt:GRCA_NL_EV_ME}  leads to the following negative log-likelihood (referred to as data fidelity term in what follows, and defined up to a multiplicative constant which is the noise variance)
\begin{eqnarray}
\mathcal{L}_{\bsP} \left(\bsZ \right)=  \frac{1}{2} ||\bsY - [\bsM,\bsP]  \bsZ  ||_F^2   
\label{eqt:data_term}
\end{eqnarray}
where $\bsY = [\bsy_1,\cdots,\bsy_N]$, $N$ is the total number of pixels, $\bsZ=\left[\bsA^{\top},\bsX^{\top}\right]^{\top}$ is 
the $(R+D)\times N$ matrix gathering the $(R\times N)$ abundance matrix $\bsA$ and the $(D \times N)$ residual coefficients $\bsX$ and $||\bsY||_F=\sqrt{\textrm{trace}\left(\bsY\bsY^{\top}\right)}$ denotes the Frobenius norm. Note that  $\bsP=\bsQ$ and $\bsx=\bgam$  (resp. $\bsP=\bsF^{\top}$ and $\bsx=\bsb$) when considering the NL model (resp. the ME model). Estimating the abundances and the residual coefficients is an ill-posed inverse problem that requires the introduction of prior knowledge (or regularization terms) about these parameters of interest. Therefore, we propose to solve the following regularized optimization problem  
\begin{eqnarray}
\mathcal{C} \left(\bsZ \right) =  \mathcal{L}_{\bsP}\left(\bsZ \right) + \textit{i}_{\mathds{R}_{+}}\left(\bsA\right) + \textit{i}_{\left\lbrace\bone_{(1,R)}\right\rbrace}\left(\bone_{(1,R)}\bsA\right)
  \nonumber \\
+  \tau_1 ||\bsX||_1 +   \tau_2 ||\bsX||_{2,1} +  \psi\left(\bsX\right)
\label{eqt:CostFunGen}
\end{eqnarray}
where $\textit{i}_{\mathds{R}_{+}}\left(\bsA\right)=\sum_{n=1}^{N}{\textit{i}_{\mathds{R}_{+}}\left(\bsa_{n}\right)}$ is the indicator function that imposes the ANC ($\textit{i}_{\mathds{R}_{+}}\left(\bsa_{n}\right)=0$  if $\bsa_{n}$ belongs to the non-negative orthant and $+\infty$ otherwise), $\textit{i}_{\left\lbrace\bone_{(1,R)}\right\rbrace}\left(\bone_{(1,R)}\bsA\right) =\sum_{n=1}^{N}{\textit{i}_{\left\lbrace1\right\rbrace}\left(\bone_{(1,R)}\bsa_{n}\right)}$ is the indicator function that imposes the ASC to each abundance vector $\bsa_{n}$, $\bone_{(i,j)}$ denotes the $i \times j$ vector of $1$s,   $\psi\left(\bsX\right) = \textit{i}_{\mathds{R}_{+}}\left(\bsX\right)$ when considering the NL model and  $\psi\left(\bsX\right) = 0$ for the ME model.  The first line of \eqref{eqt:CostFunGen} is a sum of the quadratic data fidelity term associated with the Gaussian noise statistics and two convex terms imposing the abundance constraints. The second line  of \eqref{eqt:CostFunGen}   accounts for the sparsity behavior of the residual coefficients. 
The first convex term $||\bsX||_1= \sum_{n=1}^{N}{||\bsx_n||_1}$ is an $\ell_1$ norm that promotes  element-wise sparsity on the $D \times N$ matrix $\bsX$. This behavior is illustrated in Fig. \ref{fig:Norms_Graph} (left) which shows a point-wise repartition of the active elements of $\bsX$. The second convex term  $||\bsX||_{2,1} = \sum_{n=1}^{N}{||  \bsx_{n}||_2} = \sum_{n=1}^{N}{\sqrt{\bsx_{n}^T \bsx_{n}}}$  is the $\ell_{21}$ mixed norm of $\bsX$ which promotes sparsity among the columns of $\bsX$, i.e., it promotes solutions of \eqref{eqt:CostFunGen} with a small number of nonlinear or outlier pixels. This regularization term has   received increasing interest in recent years \cite{SprechmannIEEETSP2011,FevotteTIP2015,AggarwalJSTARS2016,IordacheTGRS2014,IordacheTGRS2014b} and is known as a collaborative regularization since it uses information about the residuals in all the pixels to promote group-sparsity over the columns of $\bsX$. The effect of this mixed norm is illustrated in Fig. \ref{fig:Norms_Graph} (middle). Equation \eqref{eqt:CostFunGen} includes a combination of the $\ell_1$ norm and the $\ell_{21}$ mixed norm which leads to a slightly different effect as highlighted in Fig. \ref{fig:Norms_Graph} (right). Indeed, this combination allows for sparsity among the elements of the columns of $\bsX$. 
Finally the cost function \eqref{eqt:CostFunGen} is a sum of convex functions that is solved using the ADMM algorithm proposed in \cite{Figueiredo_TIP2010,Afonso_TIP2011} and described in the next section. 

 
\subsection{The ADMM algorithm} \label{sec:The_ADMM_algorithm} 
Consider the optimization problem  
\begin{equation}
\operatornamewithlimits{\textrm{argmin}}\limits_{\bsZ} \mathcal{C} \left(\bsZ \right)  = \operatornamewithlimits{\textrm{argmin}}\limits_{\bsZ} \sum_{j=1}^{J}{g_j \left(\bsH_{j} \bsZ\right)} \vspace{-0.25cm}
\label{eqt:Cost_FunADMM}
\end{equation} 
where $\bsZ \in\mathds{R}^{(R+D)\times N}$,  $g_j: \mathds{R}^{p_j\times N} \rightarrow \mathds{R}$ are closed, proper, convex functions, and $\bsH_{j} \in \mathds{R}^{p_j\times (R+D)}$ are arbitrary matrices. After denoting $\bsU_{j}=\bsH_{j} \bsZ  \in \mathds{R}^{p_j\times N}$ and introducing the auxiliary variable $\bsD_{j} \in\mathds{R}^{p_j\times N}$,  the authors in \cite{Figueiredo_TIP2010,Afonso_TIP2011} introduced the ADMM variant summarized in Algo. \ref{alg:ADMM_variant_for} to solve \eqref{eqt:Cost_FunADMM} using a variable splitting and an Augmented Lagrangian algorithm. This algorithm is designed to solve any sum of an $\ell_2$ norm with convex functions. Moreover, \cite[Theorem 1]{EcksteinMP1992} states that Algo. \ref{alg:ADMM_variant_for} converges when the matrix $\bsG = \left[ \sum_{j=1}^{J}{ \left(\bsH_{j}\right)^{\top} \bsH_{j} }  \right]$ has full rank, and the functions $g_j $ are closed, proper, and convex. Under these conditions, the same theorem states that, for any $\mu > 0$, if \eqref{eqt:Cost_FunADMM} has a non-empty set of solutions, then the generated sequence ${\bsZ^{(k)}}$  converges to a solution. If  \eqref{eqt:Cost_FunADMM} does not have a solution, then at least one of the sequences ${\bsU^{(k)}}$ or ${\bsD^{(k)}}$ diverges. These conditions will be studied for each of the proposed optimization problems in the next sections. 
Note that the main steps of Algo. \ref{alg:ADMM_variant_for}, in each iteration, are  the solution of a linear system of equations (line 8), the computation of the Moreau proximity operators (MPOs) \cite{Combettes2011Book} (line 12), and the updating of the Lagrange multipliers (line 16). More details regarding these computations are provided in Appendix \ref{app:ADMM_algorithm}.
Another important point to note is that the setting of $\mu$ has a strong impact on the convergence speed of the algorithm. In this paper,  $\mu$ is updated using the adaptive procedure described in \cite{IordacheTGRS2014,Boyd_FTML2011}, whose objective is to keep the ratio between the ADMM primal and dual residual norms within a given positive interval, as they both converge to zero. Note finally that the algorithms are stopped if the primal or dual residual norms are lower than a given threshold \cite{Boyd_FTML2011}.
We refer the reader to \cite{Figueiredo_TIP2010,Afonso_TIP2011,IordacheTGRS2014,Boyd_FTML2011} for more details regarding the ADMM algorithm. 

\begin{algorithm}
\caption{ADMM variant for \eqref{eqt:Cost_FunADMM}} \label{alg:ADMM_variant_for}
\begin{algorithmic}[1]
       \STATE \underline{Initialization}
       \STATE Initialize $\bsU_{j}^{(0)}, \bsD_{j}^{(0)}, \forall j$, $\mu >0$. Set $k\leftarrow 0$, conv$\leftarrow 0$    
       \WHILE{conv$=0$}
                  \FOR{j=1:J} 
							             \STATE $\xi_{j}^{(k)}\leftarrow \bsU_{j}^{(k)}+\bsD_{j}^{(k)}$,
							       \ENDFOR
							 \STATE \underline{Linear system of equations}	
							 \STATE $\bsZ^{(k+1)} \leftarrow \bsG^{-1}   \sum_{j=1}^{J}{\left(\bsH_{j}\right)^{\top} \xi_{j}^{(k)}}$,							
							\STATE \underline{Moreau proximity operators}	
							   \FOR{j=1:J} 
							             \STATE $\bsV_{j}^{(k)} \leftarrow \bsH_{j} \bsZ^{(k+1)} - \bsD_{j}^{(k)}$,
													 \STATE $\bsU_{j}^{(k+1)} \leftarrow  \operatornamewithlimits{\textrm{argmin}}\limits_{\bsU_{j}}  \frac{\mu}{2}   ||\bsU_{j} - \bsV_{j}^{(k)}||^2 + g_j\left(\bsU_{j} \right)$,
							   \ENDFOR 
								\STATE \underline{Update Lagrange multipliers}	
						     \FOR{j=1:J} 
							             \STATE $\bsD_{j}^{(k+1)}  \leftarrow \bsU_{j}^{(k+1)} - \bsV_{j}^{(k)}$,
							   \ENDFOR 
               \STATE $k = k  + 1$
       \ENDWHILE 
\end{algorithmic}
\end{algorithm}

\subsection{The NUSAL-$K$ algorithm} \label{sec:NUSAL_algorithm} 
This section presents the optimization problem considered for estimating the parameters of  the NL model \eqref{eqt:GRCA_NL}.   We first recall the two assumptions: (i) the nonlinearity appears in some pixels of the image, (ii) in a nonlinear pixel, only a few interactions are active. Under these considerations, we propose to solve the following optimization problem       
\begin{eqnarray}
\mathcal{C_{\textrm{NUSAL-$K$}}} \left(\bsZ \right) & = & \frac{1}{2} ||\bsY - [\bsM,\bsQ]  \bsZ  ||_F^2 
   \nonumber \\
& + & \tau_1 ||\bGam||_1 +   \tau_2 ||\bGam||_{2,1}  \nonumber \\ 
& + & \textit{i}_{\mathds{R}_{+}}\left(\bsZ\right) + \textit{i}_{\left\lbrace\bone_{(1,R)}\right\rbrace}\left(\bone_{(1,R)}\bsA\right)   
\label{eqt:CostFun_NUSAL}
\end{eqnarray}  
where $\bGam=\left[\bgam_{1},\cdots,\bgam_{N}\right]$ is a $(D_{K}\times N)$ matrix of nonlinear coefficients, and $\bsZ=\left[\bsA^{\top},\bGam^{\top}\right]^{\top}$. The mixed norm $\ell_{21}$ imposes sparsity on the nonlinear pixels, i.e., it imposes sparsity on the columns of  $\bGam$ (see Fig. \ref{fig:Norms_Graph}). In addition, the $\ell_{1}$ norm further enforces sparsity on the nonlinear interactions in the active nonlinear pixels as highlighted in Fig. \ref{fig:Norms_Graph} (right). Using the same notation as in \eqref{eqt:Cost_FunADMM}, problem \eqref{eqt:CostFun_NUSAL} can be expressed as the sum of $J=5$ convex terms given by 
\begin{equation}
     \begin{aligned} 
g_1\left(\bsU_{1}\right) = & \, \mathcal{L}_{\bsQ} \left(\bsU_{1}\right),& \; &\bsH_{1} = \mathds{I}_{(R+D_K)}   \\ 
g_2\left(\bsU_{2}\right) = &  \, \tau_1 {||\bsU_{2}||_1},& \; &\bsH_{2} =\left[\bzero_{(D_K,R)}, \mathds{I}_{D_K} \right]  \\  
g_3\left(\bsU_{3}\right) = & \, \tau_2 {||\bsU_{3}||_{2,1}},& \; &\bsH_{3} =\left[\bzero_{(D_K,R)}, \mathds{I}_{D_K} \right]  \\ 
g_4\left(\bsU_{4}\right) = & \, \textit{i}_{\mathds{R}_{+}}\left(\bsU_{4}\right),& \; &\bsH_{4} =\mathds{I}_{(R+D_K)}  \\ 
g_5\left(\bsU_{5}\right) = & \, \textit{i}_{\left\lbrace\bone^{\top}\right\rbrace}\left(\bone^{\top}\bsU_{5} \right),& \; &\bsH_{5} =\left[\mathds{I}_{R}, \bzero_{(R,D_K)}  \right]  \\
\end{aligned}  
\label{eqt:Def_gi_Hi_NUSAL} 
\end{equation} 
where $\mathds{I}_{n}$ denotes the $n \times n$ identity matrix and $\bzero_{(i,j)}$ denotes the $i \times j$ matrix of zeros. For this problem, the matrix $\bsG$ is given by $\bsG = \diag{[3\bone_{(1,R)},4\bone_{(1,D_K)}]}$ which is clearly of full rank. This matrix and the properties of $g_i, i \in\left\lbrace 1,\cdots,J \right\rbrace$ ensures the algorithm convergence.
 
\subsection{The RUSAL algorithm} \label{sec:RUSAL_algorithm} 
The optimization problem used to estimate the parameters of the ME model \eqref{eqt:GRCA_RL}  is based on following assumptions: (i) the outliers appear at some pixels of the image, (ii) the residual spectra are smooth (i.e., the DCT coefficients are sparse). Under these considerations, we propose to solve the following optimization problem   
\begin{eqnarray}
\mathcal{C_{\textrm{RUSAL}}} \left(\bsZ \right) & = & \frac{1}{2} ||\bsY - [\bsM,\bsF^T]  \bsZ  ||_F^2 
   \nonumber \\
& + &  \tau_1 ||\bsB||_1 +   \tau_2 ||\bsB||_{2,1}  \nonumber \\ 
& + & \textit{i}_{\mathds{R}_{+}}\left(\bsA\right) + \textit{i}_{\left\lbrace\bone_{(1,R)}\right\rbrace}\left(\bone_{(1,R)}\bsA\right)   
\label{eqt:CostFunRUSAL}
\end{eqnarray}   
where $\bsB=\left[\bsb_{1},\cdots,\bsb_{N}\right]$, and $\bsZ=\left[\bsA^{\top},\bsB^{\top}\right]^{\top}$. In a similar fashion to NUSAL-$K$, the mixed norm $\ell_{21}$ ensures spatial sparsity of the mismodelling coefficients  $\bsB$. In addition, the $\ell_{1}$ norm further enforces sparsity on the DCT coefficients of each active pixel to impose spectral smoothness of the residuals. Using the same notation as in \eqref{eqt:Cost_FunADMM}, problem \eqref{eqt:CostFunRUSAL} can be expressed as the sum of $J=5$ convex terms given by  
\begin{equation}
     \begin{aligned}  
g_1\left(\bsU_{1}\right) = & \, \mathcal{L}_{\bsF^{\top}} \left(\bsU_{1}\right),& \; & \bsH_{1} = \mathds{I}_{(R+D)}   \\ 
g_2\left(\bsU_{2}\right) = & \, \tau_1 {||\bsU_{2}||_1},& \; & \bsH_{2} =\left[\bzero_{(D,R)}, \mathds{I}_{D} \right]  \\  
g_3\left(\bsU_{3}\right) = & \, \tau_2 {||\bsU_{3}||_{2,1}},& \; & \bsH_{3} =\left[\bzero_{(D,R)}, \mathds{I}_{D} \right]  \\ 
g_4\left(\bsU_{4}\right) = & \, \textit{i}_{\mathds{R}_{+}}\left(\bsU_{4}\right),& \; & \bsH_{4} =\left[\mathds{I}_{R}, \bzero_{(R,D)}  \right] \\ 
g_5\left(\bsU_{5}\right) = & \, \textit{i}_{\left\lbrace\bone^{\top}\right\rbrace}\left(\bone^{\top}\bsU_{5} \right),& \; & \bsH_{5} =\left[\mathds{I}_{R}, \bzero_{(R,D)}  \right].  \\
\end{aligned}  
\label{eqt:Def_gi_Hi_RUSAL}
\end{equation} 
For this problem, the  full rank matrix $\bsG$ is given by $\bsG = 3\mathds{I}_{(R+D)}$ which, in addition to the properties of $g_i, i \in\left\lbrace 1,\cdots,J \right\rbrace$, ensures the algorithm convergence.
 
\subsection{Computational complexity} \label{sec:Computational_complexity}  
The ADMM algorithm involves the iterative update of the matrices  $\bsZ\in\mathds{R}^{(R+D)\times N}$ (line 8 in algo. \ref{alg:ADMM_variant_for}) and $\bsU \in \mathds{R}^{p_j\times N}$ (line 12 in algo. \ref{alg:ADMM_variant_for}), where the details of the optimizations with respect to $\bsU_j$, $j\in \left\lbrace 1,\cdots,5 \right\rbrace$   are provided in the Appendix \ref{app:ADMM_algorithm}. The computational complexity of Algo. \ref{alg:ADMM_variant_for} per iteration is  $\mathcal{O}\left((R+D)^2 N \right)$, which is related to the most expensive step introduced by the calculus of $\bsU_{1}$. Finally, it is interesting to note that the matrices to inverse involve low complexity since the matrix $\bsG$ in line 8 is diagonal, and the matrix to inverse to update  $\bsU_{1}$  is fixed and then can be precomputed outside the iterative loop.

\section{Simulation results on synthetic data} \label{sec:Simulation_results_on_synthetic_data}
This section evaluates the performance of the proposed algorithms with synthetic data. This enables the performance of the algorithms to be compared on data with a known ground truth. All simulations have been implemented using MATLAB R2015a  on a computer with Intel(R) Core(TM) i7-4790 CPU@3.60GHz and 32GB RAM. The section is divided into three parts whose objectives are: 1) introducing the criteria used for the evaluation of the unmixing quality, 2) description of the synthetic images considered in the experiments, and 3) evaluating and comparing the proposed NUSAL-$K$ and RUSAL algorithms with other state-of-the-art algorithms.   

\subsection{Evaluation criteria} \label{subsec:Evaluation_criteria}
The performance of the algorithm has been assessed in terms of abundance estimation by comparing the estimated and actual abundances using the average root mean square error (aRMSE) defined by
$\textrm{aRMSE}\left(\bsA\right) = \sqrt{\frac{1}{N\, R}\sum_{n=1}^{N}
\left\| \bsa_n-\hat{\bsa}_n \right\|_2^{2}}$. As a measure of fit, we consider the following reconstruction error  $\textrm{RE}= \sqrt{\frac{1}{N\,L}  \sum_{n=1}^{N} \left\| \hat{\bsy}_n-\bsy_n  \right\|_2^2}$ and   spectral angle mapper $\textrm{SAM}  = \frac{1}{N}  \sum_{n=1}^{N} \arccos \left(\frac{\hat{\bsy}_n^T \bsy_n}{\left\|\bsy_n\right\|_2 \; \left\|\hat{\bsy}_n  \right\|_2}\right)$ criteria, where $\arccos(\cdot)$ is the inverse cosine operator and $\bsy_n$, $\hat{\bsy}_n$ denotes  the $\#n$th measured and estimated pixel spectra.

\subsection{Description of the synthetic images} \label{subsec:Description_of_the_synthetic_images} 
\begin{figure} [h!]
\centering
\includegraphics[width=0.95\figwidth]{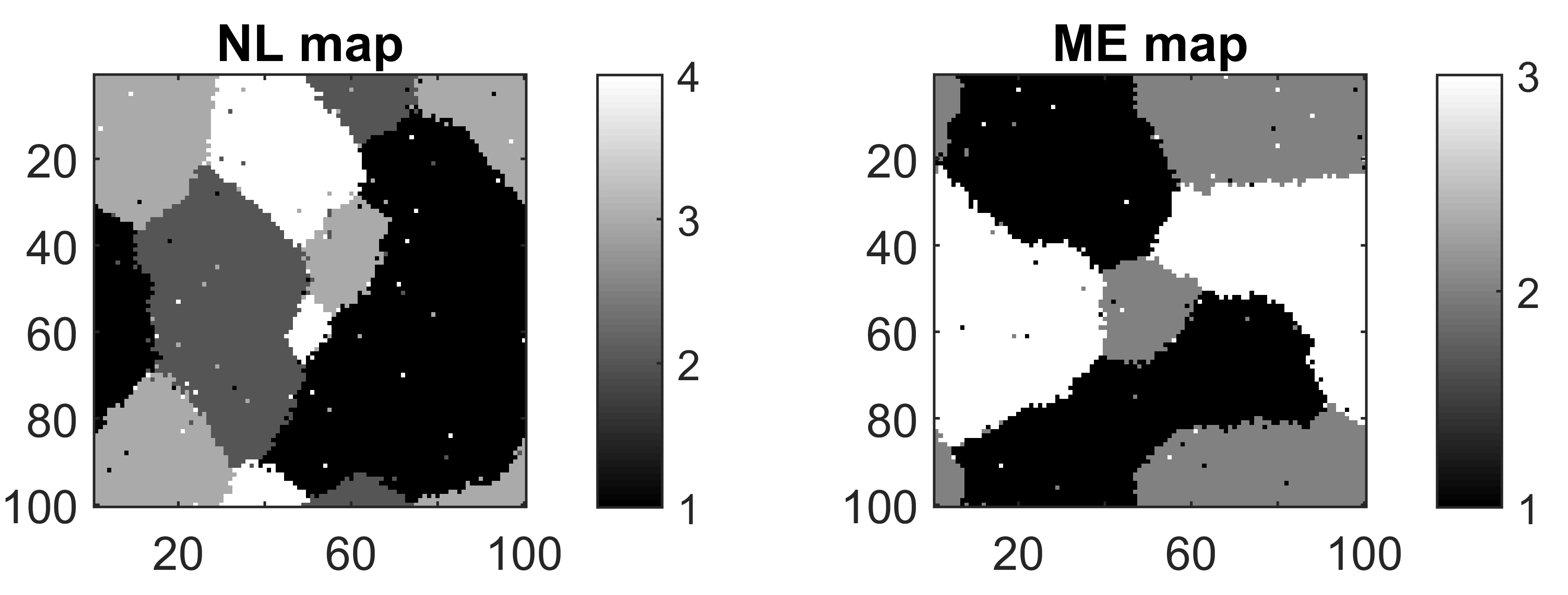}
\caption{Label maps associated with (left) the NL synthetic image, (right) the ME synthetic image. } \label{fig:Synth_Maps_NL_ME}
\end{figure}

The proposed unmixing algorithms are evaluated on two images with different parameters. The images of size  $100 \times 100$ pixels and $L=207$ spectral bands have been generated using $R$ endmembers corresponding to spectral signatures available in the ENVI software library \cite{ENVImanual2003}. All images have been corrupted by  i.i.d. Gaussian noise of variance $\sigma^2$ whose level is adjusted to obtain SNR$=25$ dB where $\textrm{SNR} = 10 \log{\left(\frac{||\bsM \bsA + \bPhi||_F^2}{LN \sigma^2}\right)}$. The images have been generated using different mixture models as follows
\begin{itemize} 
\item Linear+Nonlinear models:  image $I_1$ has been generated with $4$ linear/nonlinear models. An image partition into $4$ classes has been generated by considering a Potts-Markov random field (with granularity parameter $\beta= 0.8$) as shown in Fig. \ref{fig:Synth_Maps_NL_ME} (left). The four spatial classes are associated with the LMM, NL-$3$ model \eqref{eqt:GRCA_NL} (with $\bgam_{n}  \sim \calN_{{(\dsR+)}^D} \left(\bzero_{D,1}, 0.1 \mathds{I}_D \right)$),  GBM (with random nonlinear coefficients in $[0.8,1]$) and PPNMM (with $b=0.5$), respectively. Note that the generated nonlinear coefficients $\bgam_{n}$ are not sparse, which is a challenging scenario for the NUSAL-$K$ algorithm. Finally, the abundances have been generated uniformly in the simplex of ANC and ASC. 
\item Mismodelling effects:  image $I_2$ has been partitioned into $3$ classes by considering a Potts-Markov random field  as shown in Fig. \ref{fig:Synth_Maps_NL_ME} (right).  Pixels of the first class have been generated according to the LMM model, and  the pixels of class $2$ have been generated while considering EV. This has been achieved by varying the endmembers in each pixel of the image.  Indeed, a pixel dependent smooth spectral function $\bsp_{rn} \in \mathds{R}^{L\times 1}$ has been added to each endmember to model EV. As in \cite{HalimiTIP2016}, the smooth functions were generated as follows $\bsp_{rn} \sim \mathcal{N} (\bzero_{L\times 1}, \epsilon^2 \bSig_{\bsp})$, where $\bSig_{\bsp}$ is an $\left(L\times L\right)$ squared-exponential covariance matrix modeling the spectral correlations and $\epsilon^2=0.001$.   The pixels of class $3$ have been generated according to the ME model proposed in \cite{HalimiTIP2016},  since it leads to smooth spectral residuals as in \eqref{eqt:GRCA_RL}. More precisely, the residuals have been generated as follows $\bphi_{n}^{\textrm{ME}} \sim \mathcal{N} (\bzero_{L\times 1}, \epsilon^2 \bSig_{\bsp})$, with $\epsilon^2=0.002$. Finally, the abundances have been generated uniformly in the simplex of ANC and ASC.
\end{itemize}
Note that both images have been generated with the number of endmembers varying in the interval $\left\lbrace 3,6 \right\rbrace$. 
 
\setlength{\tabcolsep}{5pt}
\begin{table*} \centering
\centering \caption{Results on the LMM-NL based synthetic image $I_1$ for $R \in \left\lbrace 3,6 \right\rbrace$  endmembers and SNR$=25$ dB.}
{\tiny
\begin{tabular}{|c|c|c|c|c|c|c|c|c|||c|c|c|c|c|c|c|c|}
	\cline{2-17}
\multicolumn{1}{c|}{}                & \multicolumn{8}{c|||}{R$=3$}     & \multicolumn{8}{c|}{R$=6$ } \\
\cline{2-17}
\multicolumn{1}{c|}{}                & \multicolumn{4}{c|}{RMSE}     &  \multirow{3}{*}{RMSE}    &  \multirow{3}{*}{RE}   &  \multirow{3}{*}{SAM}  & \multirow{3}{*}{Time}  & \multicolumn{4}{c|}{RMSE}     & \multirow{3}{*}{RMSE}    &  \multirow{3}{*}{RE}   &  \multirow{3}{*}{SAM}  & \multirow{3}{*}{Time}  \\
\cline{2-5}\cline{10-14}
\multicolumn{1}{c|}{}                & $\mathcal{C}_1$  & $\mathcal{C}_2$ & $\mathcal{C}_3$ & $\mathcal{C}_4$     &    &     &    &    & $\mathcal{C}_1$  & $\mathcal{C}_2$ & $\mathcal{C}_3$ & $\mathcal{C}_4$     &   &     &    &  \\
\multicolumn{1}{c|}{}                &  LMM  & NL-$3$ & GBM & PPNMM     &      &      &     &   & LMM  & NL-$3$ & GBM & PPNMM      &      &      &  &    \\
\hline   FCLS              & \red{$1.4$} & $20.3$ & $5.8$ & $11.8$ & $10.82$ & $9.66$ & $7.62$ & \red{$  1$}   & \red{$3.7$} & $39.8$ & $13.0$ & $16.8$   & $20.78$ & $27.72$ & $11.42$ &   \blue{$2$} \\
\hline   SUNSAL            & \red{$1.4$} & $20.3$ & $5.8$ & $11.9$ & $10.82$ & $9.66$ & $7.62$ & \blue{$  0.1$}   & \red{$3.7$} & $40.6$ & $13.0$ & $16.8$   & $21.11$ & $27.26$ & $11.42$ & \red{$0.2$} \\
\hline  SKhype             &  $2.2$ & $11.7$ & $3.0$ & \red{$3.9$} & $6.01$ & $-$ & $-$ & $466$   & $5.0$ & $16.7$ & \blue{$5.4$} & \blue{$5.7$}   & $9.09$ & $-$ & $-$ & $201$  \\
\hline  CDA-NL             &  \red{$1.4$} & $4.5$ & \blue{$2.1$} & \blue{$4.2$}  & $2.93$ & $2.67$ & $5.79$ & $182$   & \red{$3.7$} & $13.2$ & $11.5$ & $6.1$   & $9.00$ & $3.50$ & $7.64$ & $1560$   \\
\hline  CDA-EV             & $3.3$ & $23.3$ & $4.5$ & $9.6$  & $11.84$ & $2.94$ & $6.18$ & $246$   & $5.0$ & $37.5$ & $6.7$ & $8.6$   & $18.42$ & $6.56$ & $9.17$ & $555$  \\
\hline  CDA-ME             & $1.8$ & $21.2$ & $5.4$ & $11.1$ & $11.04$ & \red{$2.59$} & \red{$5.66$} & $ 64$   & $4.1$ & $30.9$ & $7.9$ & $9.6$   & $15.63$ & \blue{$3.14$} & \blue{$7.17$} & $ 66$ \\
\hline   RNMF              & $1.5$ & $12.8$ & $2.5$ & $5.2$  & $6.45$ & $8.04$ & $6.77$ & $110$   & $5.3$ & $22.2$ & $6.7$ & $8.1$    & $11.79$ & \red{$2.63$} & \red{$3.85$} & $ 50$   \\
\hline  \textbf{RUSAL}     & \red{$1.4$} & $17.8$ & $6.5$ & $10.6$ & $9.72$ & $3.04$ & $6.24$ & $ 48$   & $5.7$ & $38.0$ & $11.7$ & $14.5$  & $19.74$ & $3.54$ & $7.37$ & $ 35$ \\
\hline  \textbf{NUSAL-$2$}   & \red{$1.4$} & \blue{$3.9$} & \red{$2.0$} & $5.0$    & \blue{$2.88$} & $2.69$ & $5.81$ & $  7$   & \red{$3.7$} & \blue{$9.1$} & $5.7$ & $6.2$     & \blue{$6.04$} & $3.42$ & $7.47$ & $ 26$  \\
\hline  \textbf{NUSAL-$3$}   & \red{$1.4$} & \red{$2.9$} & \red{$2.0$} & $4.9$    & \red{$2.59$} & \blue{$2.65$} & \blue{$5.75$} & $ 19$   & \red{$3.7$} & \red{$7.4$} & \red{$4.6$} & \red{$5.4$}     & \red{$5.16$} & $3.25$ & $7.29$ & $ 96$ \\
\hline
\end{tabular}}
\label{tab:Results_Synth_NL_LMM_25dB}
\end{table*}

\begin{table*} \centering
\centering \caption{Results on the LMM-ME based synthetic image $I_2$ for $R \in \left\lbrace 3,6 \right\rbrace$  endmembers and SNR$=25$ dB.}
\begin{tabular}{|c|c|c|c|c|c|c|c|||c|c|c|c|c|c|c|}
	\cline{2-15}
\multicolumn{1}{c|}{}                & \multicolumn{7}{c|||}{R$=3$}     & \multicolumn{7}{c|}{R$=6$ } \\
\cline{2-15}
\multicolumn{1}{c|}{}                & \multicolumn{3}{c|}{RMSE}     &  \multirow{3}{*}{RMSE}    &  \multirow{3}{*}{RE}   &  \multirow{3}{*}{SAM}  & \multirow{3}{*}{Time}  & \multicolumn{3}{c|}{RMSE}     & \multirow{3}{*}{RMSE}    &  \multirow{3}{*}{RE}   &  \multirow{3}{*}{SAM}  & \multirow{3}{*}{Time}  \\
\cline{2-4}\cline{9-12}
\multicolumn{1}{c|}{}                & $\mathcal{C}_1$  & $\mathcal{C}_2$ & $\mathcal{C}_3$  &    &     &    &    & $\mathcal{C}_1$  & $\mathcal{C}_2$ & $\mathcal{C}_3$ &    &     &    &  \\
\multicolumn{1}{c|}{}                &  LMM  & EV & RCA      &      &      &     &   & LMM  & EV & RCA      &      &      &  &    \\
\hline   FCLS              &  \red{$1.2$} & $6.7$   & $12.6$   & $8.1$ & $3.3$ & $5.7$ & \blue{$  1$}   & \red{$2.4$} & $6.8$ &  $13.4$   & $8.6$ & $2.5$ & $5.3$ & \blue{$  1$}    \\
\hline   SUNSAL            &  \red{$1.2$} & $6.7$   & $12.6$   & $8.1$ & $3.3$ & $5.7$ & \red{$0.1$}   & \red{$2.4$} & $6.8$ &  $13.4$   & $8.6$ & $2.5$ & $5.3$ & \red{$  0.2$}   \\
\hline  SKhype             &  $1.9$ & $5.2$   & $9.2$    & $6.0$ & $-$ & $-$ & $437$   & $3.2$ & $4.7$ &  \blue{$8.4$}    & \blue{$5.8$} & $-$ & $-$ & $223$   \\
\hline  CDA-NL             &  \red{$1.2$} & $6.9$   & $13.2$   & $8.4$ & $2.7$ & $5.7$ & $317$   & \red{$2.4$} & $6.6$ &  $12.3$   & $8.0$ & $2.2$ & $5.0$ & $2278$   \\
\hline  CDA-EV             &  $2.3$ & \blue{$3.5$}   & \blue{$6.8$}    & \blue{$4.5$} & \blue{$2.3$} & \blue{$5.0$} & $227$   & $3.0$ & \red{$4.6$} &  $9.9$    & $6.5$ & \red{$2.0$} & $4.7$ & $391$   \\
\hline  CDA-ME             &  $1.5$ & \red{$3.3$}   & \red{$3.8$}    & \red{$2.9$} & \blue{$2.3$} & \blue{$5.0$} & $ 49$   & $2.7$ & \red{$4.6$} &  \red{$6.0$}    & \red{$4.5$} & \red{$2.0$} & $4.7$ & $ 71$   \\
\hline   RNMF              &  \red{$1.2$} & $6.8$   & $12.9$   & $8.3$ & $3.4$ & $5.7$ & $137$   & $2.6$ & $5.8$ &  $12.2$   & $7.8$ & $2.1$ & \red{$4.5$} & $279$  \\
\hline  \textbf{RUSAL}     &  $1.3$ & $5.4$   & $8.9$    & $5.9$ & \red{$2.2$} & \red{$4.8$} & $ 31$   & $2.5$ & $6.1$ &  $11.1$   & $7.2$ & \red{$2.0$} & \blue{$4.6$} & $ 35$  \\
\hline  \textbf{NUSAL-$2$}    &  \red{$1.2$} & $6.8$   & $12.8$   & $8.1$ & $3.0$ & $5.7$ & $  6$   & \red{$2.4$} & $6.1$ &  $12.8$   & $8.1$ & $2.3$ & $5.1$ & $ 41$   \\
\hline  \textbf{NUSAL-$3$}    &  \red{$1.2$} & $6.8$   & $12.8$   & $8.2$ & $3.0$ & $5.7$ & $ 19$   & $2.5$ & $6.1$ &  $12.8$   & $8.1$ & $2.3$ & $5.1$ & $138$ \\
\hline
\end{tabular}
\label{tab:Results_Synth_ME_LMM_25dB}
\end{table*}

\subsection{Performance of the proposed algorithms} \label{sec:Performance_of_the_proposed_algorithms}
The proposed RUSAL and NUSAL-$K$ algorithms are compared to state-of-the-art algorithms by processing the generated synthetic images. We consider the two variants NUSAL-$2$ and  NUSAL-$3$ to study the effect of high order interaction terms. The comparison algorithms are associated with different mixture models as follows
\begin{itemize}
\item Linear unmixing: the abundances are estimated using the FCLS algorithm \cite{Heinz2001} and the SUNSAL algorithm \cite{BioucasWhispers2010}.
\item Nonlinear unmixing: the abundances are estimated using the CDA-NL algorithm  \cite{HalimiTIP2016} and the SKhype algorithm \cite{ChenTSP2013}
\item Endmember variability: the abundances are estimated using the CDA-EV algorithm  \cite{HalimiTIP2016}
\item Mismodelling effects (robust algorithms): the abundances are estimated using the CDA-ME algorithm  \cite{HalimiTIP2016} and the RNMF algorithm\footnote{The RNMF was introduced in \cite{FevotteTIP2015} as a nonlinear algorithm. In this paper, we consider it as an intermediate model between ME models (since it does not account for multiple scattering) and NL model (since it includes the non-negativity constraint). } 
\cite{FevotteTIP2015}.
\end{itemize}
For comparison purposes,  the endmembers of these algorithms have been fixed to the actual spectra used to generate the data (the endmember update step in RNMF has been removed). Moreover, the CDA algorithms have been used while fixing the illumination coefficient to the value $\#1$ to provide a fair comparison with the remaining algorithms. Note also that the RNMF, NUSAL-$K$ and RUSAL algorithms require the regularization parameters to be set. In this study, we provide the best performance (in terms of abundance RMSE) of these algorithms when varying the regularization parameters as follows: $\lambda$ of RNMF varies in $ \left\lbrace 0.01\lambda_0, 0.1\lambda_0,  \lambda_0  \right\rbrace$ (where $\lambda_0$ has been suggested in \cite{FevotteTIP2015}), for RUSAL:  $\tau_1$ and $\tau_2$ vary in $\left\lbrace 0.001, 0.003, 0.006, 0.01, 0.05, 0.1 \right\rbrace$ and $D=20$ in all experiments, and for NUSAL-$K$  $\tau_1$ and $\tau_2$ vary in $\left\lbrace 0.01, 0.05, 0.1 \right\rbrace$.  
Table \ref{tab:Results_Synth_NL_LMM_25dB} reports the results when processing the first image with $R \in \left\lbrace 3,6 \right\rbrace$ endmembers.  The RMSE of each spatial class (associated with different mixture model) are also reported. The proposed NUSAL-$2$ and NUSAL-$3$ algorithms provide the best RMSE performance for the LMM, RCA-NL-$3$ and the GBM pixels. For PPNMM, the best RMSE is obtained with SKhype that is well adapted to this polynomial nonlinearity. The best overall RMSE is obtained by the NUSAL-$2$ and NUSAL-$3$ algorithms with a slightly better values for NUSAL-$3$ since it estimates more parameters than NUSAL-$2$. Except for the LMM-based algorithms, the data are well fitted by the algorithms as indicated by the values of RE and SAM. Moreover, it is important to mention the reduced computational time of the proposed NUSAL-$K$ algorithms. Indeed, Table \ref{tab:Results_Synth_NL_LMM_25dB}  clearly shows that the NUSAL-$2$ and NUSAL-$3$ algorithms are faster than the NL state-of-the-art algorithms, i.e., CDA-NL and SKhype. It is also shown that NUSAL-$3$ requires more computational time than NUSAL-$2$, while it performs slightly better. This highlights the effect of the third order nonlinear interaction terms that improve the unmixing at a price of a higher computational time. As expected,  the mismodelling-based algorithms CDA-ME, RNMF and RUSAL provide an intermediate performance between the LMM algorithms and the NL algorithms. Indeed, these algorithms are designed to deal with different effects including the NL effect. Moreover, RUSAL is less sensitive to the variation of the endmember number $R$ than NUSAL-$K$ and CDA-NL. Indeed, the latter algorithms account for interaction terms whose number increases with $R$, while RUSAL use a flexible residual formulation that is not related to $R$ (it simply accounts for the spectral smoothness of the residuals). 
Table \ref{tab:Results_Synth_ME_LMM_25dB} shows the obtained results when processing the second image that includes pixels with LMM,  EV and ME. The best RMSE performance are generally obtained with the CDA-ME algorithm. The proposed RUSAL algorithm  provides competitive abundance RMSE with RNMF  with the advantage of a reduced computational time.  In contrast with SKhype which demonstrates robust behavior, the NL based algorithms are not well adapted to these data and provide a lower unmixing quality than the ME algorithms. These results highlight the benefit of the NUSAL-$K$ and RUSAL algorithms that show competitive results when compared to the other algorithms. Moreover, the proposed algorithms exhibit a reduced computational cost that is suitable for real world applications. While both NUSAL-$K$ and CDA-NL algorithms are sensitive to an increase in the number of endmembers,  this effect is more important for CDA-NL while it is reduced for the NUSAL-$K$ algorithms that remain faster than SKhype for $R=6$ and  $K=3$ (i.e., $77$ interaction terms as reported in Table \ref{tab:D_k_examples}). Note finally that additional experiments, conducted with SNR$=15$ dB, show a reduction of the unmixing quality for all algorithms. However, the algorithms relative behavior is similar to the studied case, and the conclusions remain valid. These results are not provided here for brevity.

\section{Results on real data} \label{sec:results_on_real_data}
\begin{figure}[h!]
\centering \subfigure[Moffett image.]{\includegraphics[width=0.46\figwidth,height=4cm]{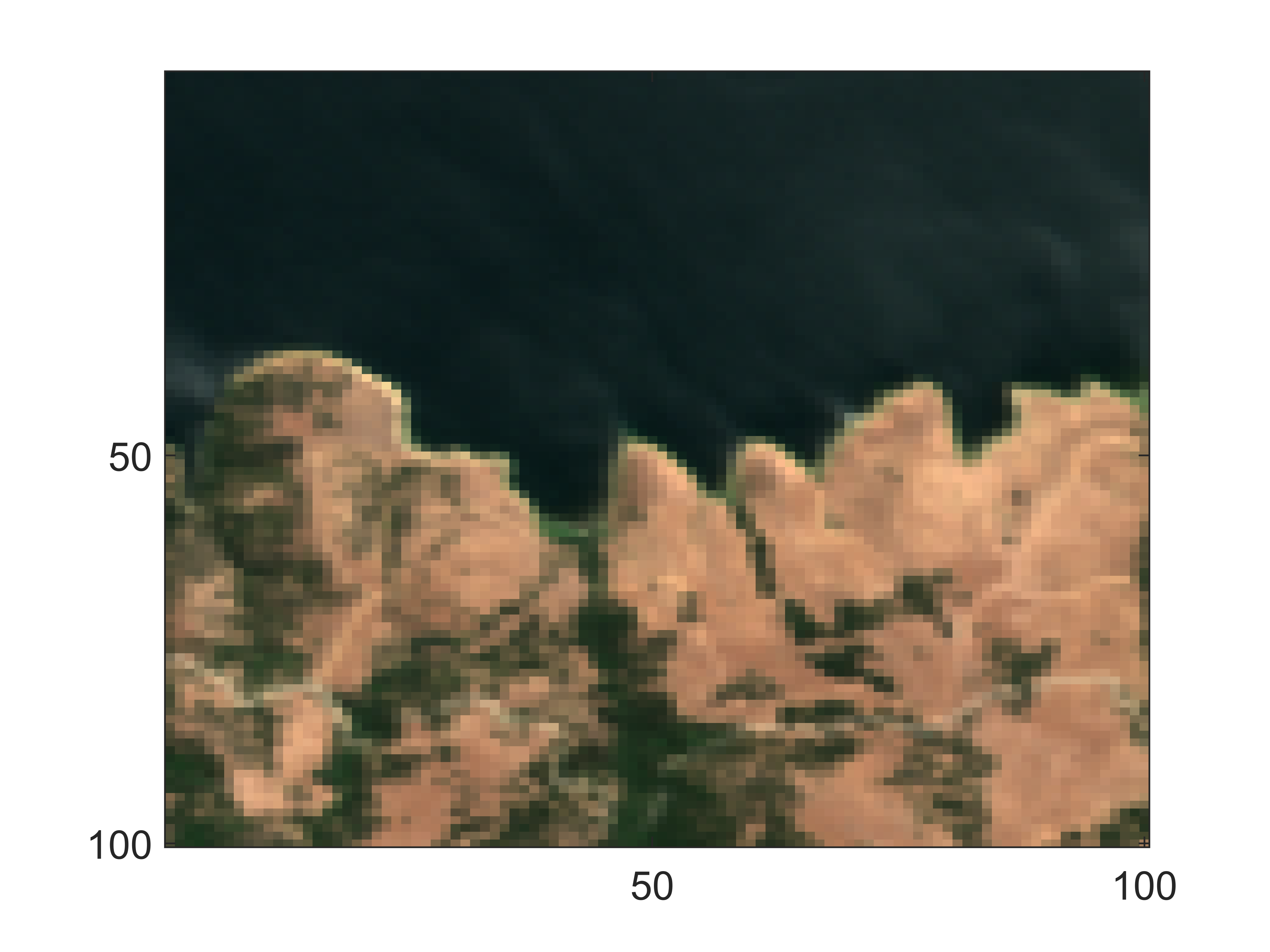}}\hspace{0.25cm}
\subfigure[Madonna image.]{\includegraphics[width=0.46\figwidth,height=4cm]{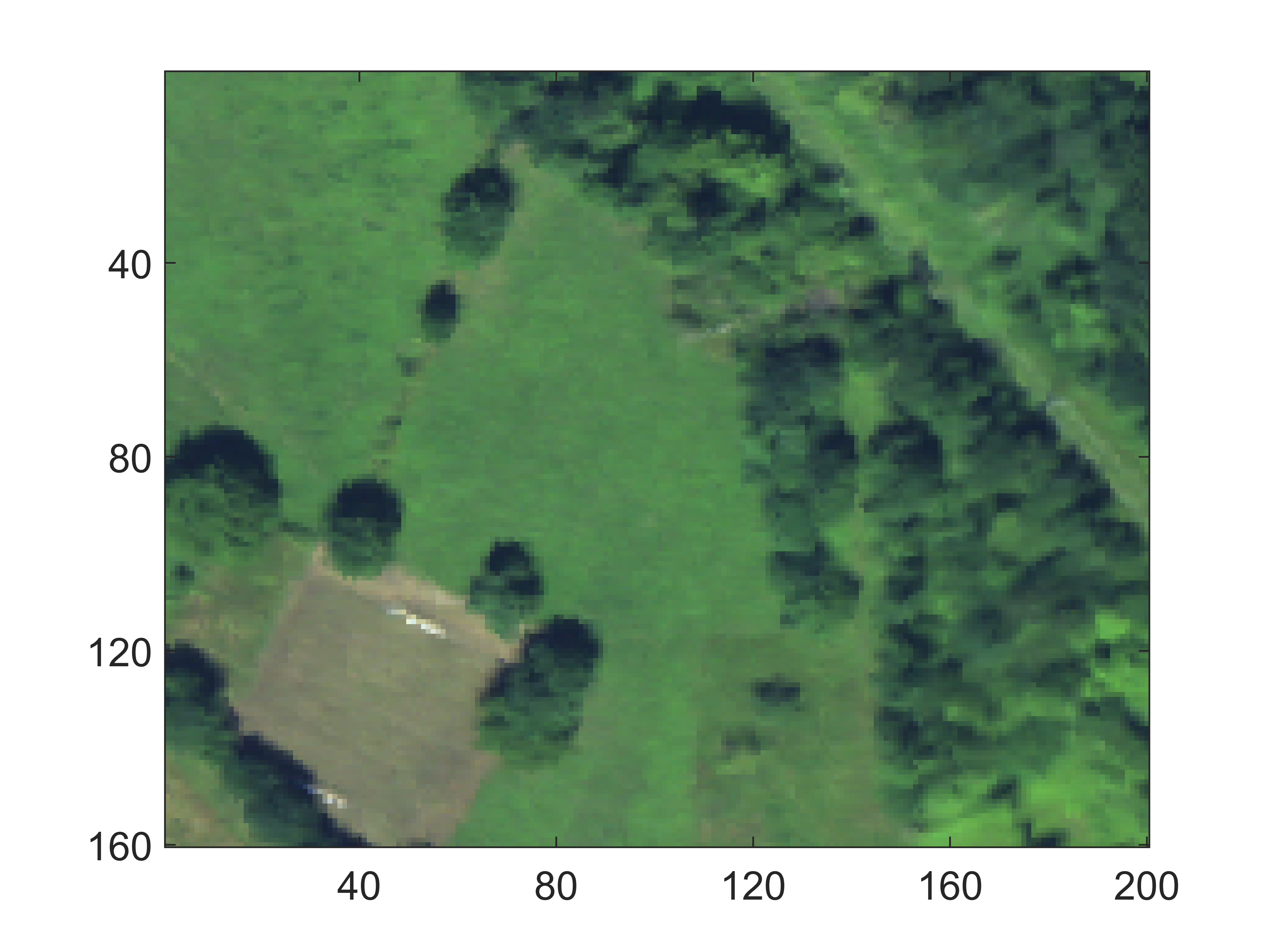}}
\subfigure[Salinas image.]{\includegraphics[width=0.46\figwidth,height=4cm]{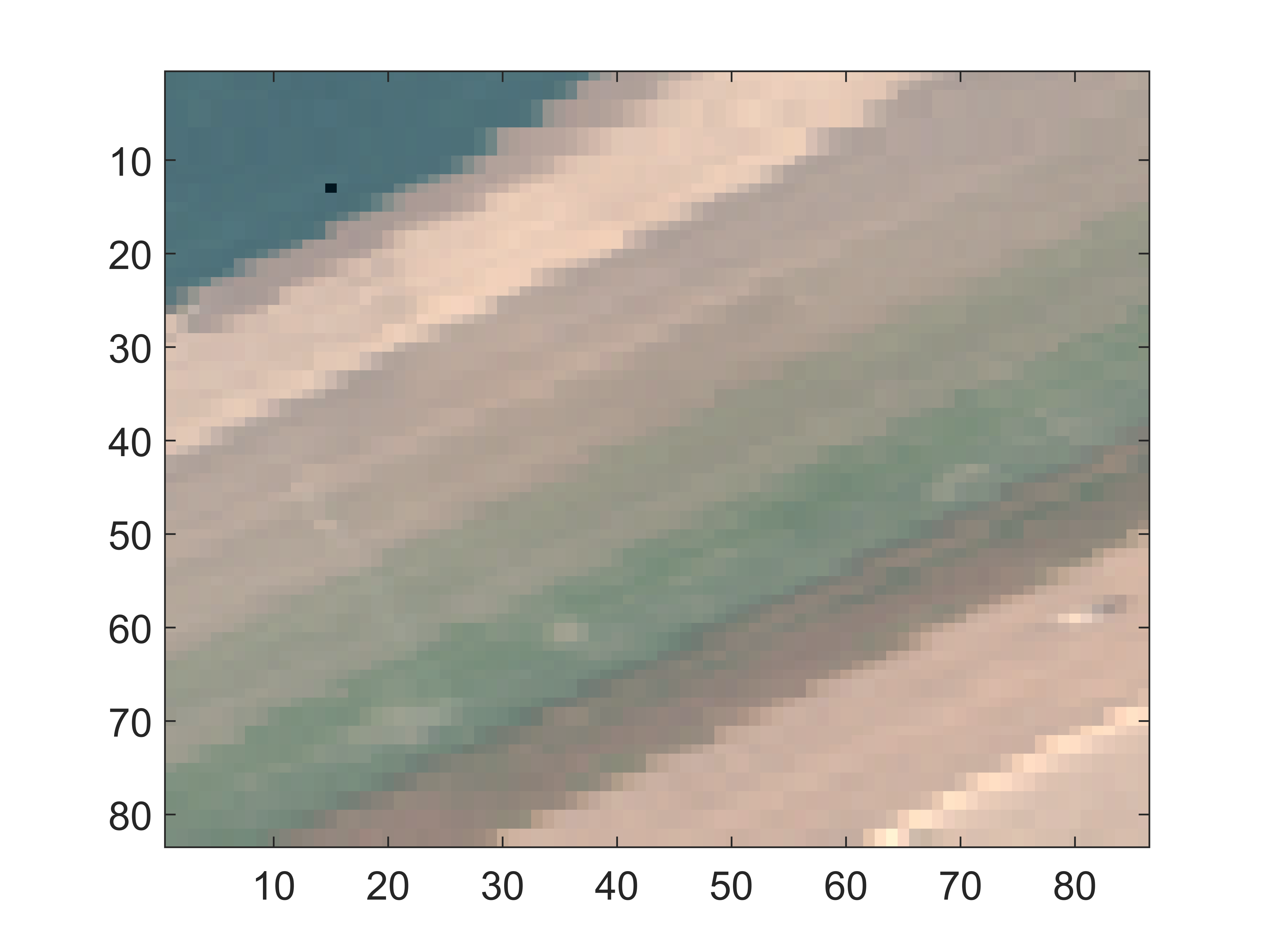}} 
\caption{Real hyperspectral images. (a) Moffett image, (b) Madonna image, (c) Salinas image.} \label{fig:TIP_Real_images}
\end{figure}

This section illustrates the performance of the proposed algorithms when applied to three real hyperspectral images. The first hyperspectral image has received much attention
in the remote sensing community \cite{Halimi2011TGRS,Dobigeon2008}. This image was acquired over Moffett Field, CA, in $1997$ by AVIRIS. The dataset contains $100 \times 100$ pixels, $L=152$ spectral bands (after removing water absorption bands) acquired in the interval $0.4-2.5 \mu$m, has a spatial resolution of $100$m and is mainly composed of three components: water, soil, and vegetation (see Fig. \ref{fig:TIP_Real_images} (a)). This image is interesting since it is known to include bilinear scattering effects \cite{Halimi2011TGRS,HalimiTIP2016,FevotteTIP2015} which makes it suitable for the assessment of the NUSAL-$K$ and RUSAL algorithms presented in this paper.  
The second  image, denoted as Madonna, was acquired in $2010$ by the Hyspex hyperspectral scanner over Villelongue, France (00 03'W and 4257'N). The dataset contains $L=160$ spectral bands recorded from the visible to near infrared ($400-1000$nm) with a spatial resolution of $0.5$m  \cite{Sheeren2011}. It has previously been studied in \cite{Halimi_TIP2015,Altmann2014b,HalimiTIP2016} showing  NL effects (between the trees and the soil), EV effects (mainly for the vegetation) and shadow effect. The subimage considered contains $160 \times 200$ pixels and is composed of $R=4$ components: tree, grass, soil and shadow (see Fig. \ref{fig:TIP_Real_images} (b)).  For these two images, the VCA algorithm \cite{Nascimento2005} was used to extract the corresponding endmembers, i.e.,  $R= 3$ endmembers for the Moffett image and $R= 4$ endmembers for the Madonna image. 
The third image  was acquired by the AVIRIS sensor, in $1998$, over Salinas Valley, California (see  Fig. \ref{fig:TIP_Real_images} (c)). The  dataset contains $86 \times 83$ pixels, $204$ spectral bands with the same spectral resolution and spectral range as the Moffett image (the water absorption bands  were removed) and a spatial resolution of $3.7$ m. This image is interesting since it includes different species of vegetables showing endmember variability, which makes it suitable for the assessment of the RUSAL algorithm. According to the ground truth information \cite{LiTGRS2010}, this image contains $6$ classes that are: Broccoli, Corn$\underline{\;}$senesced$\underline{\;}$green$\underline{\;}$weeds, lettuce of different ages (4, 5, 6, and 7 weeks). As a result of the similarity between the different spectra and the presence of highly mixed pixels \cite{PlazaTGRS2005}, we have manually extracted $4$ endmembers associated with these classes: Corn$\underline{\;}$senesced$\underline{\;}$green$\underline{\;}$weeds + lettuce-4-5, Broccoli, lettuce-6, and lettuce-7\footnote{Each endmember is obtained by averaging bundle of spectra belonging to its class.}. Indeed, these endmembers have a different shape (minimum pairwise angle of $9$ degrees) while the remaining fluctuations can be associated with the effect of EV. 

\begin{figure}[h!]
\centering
\includegraphics[width=0.95\figwidth,height=8cm]{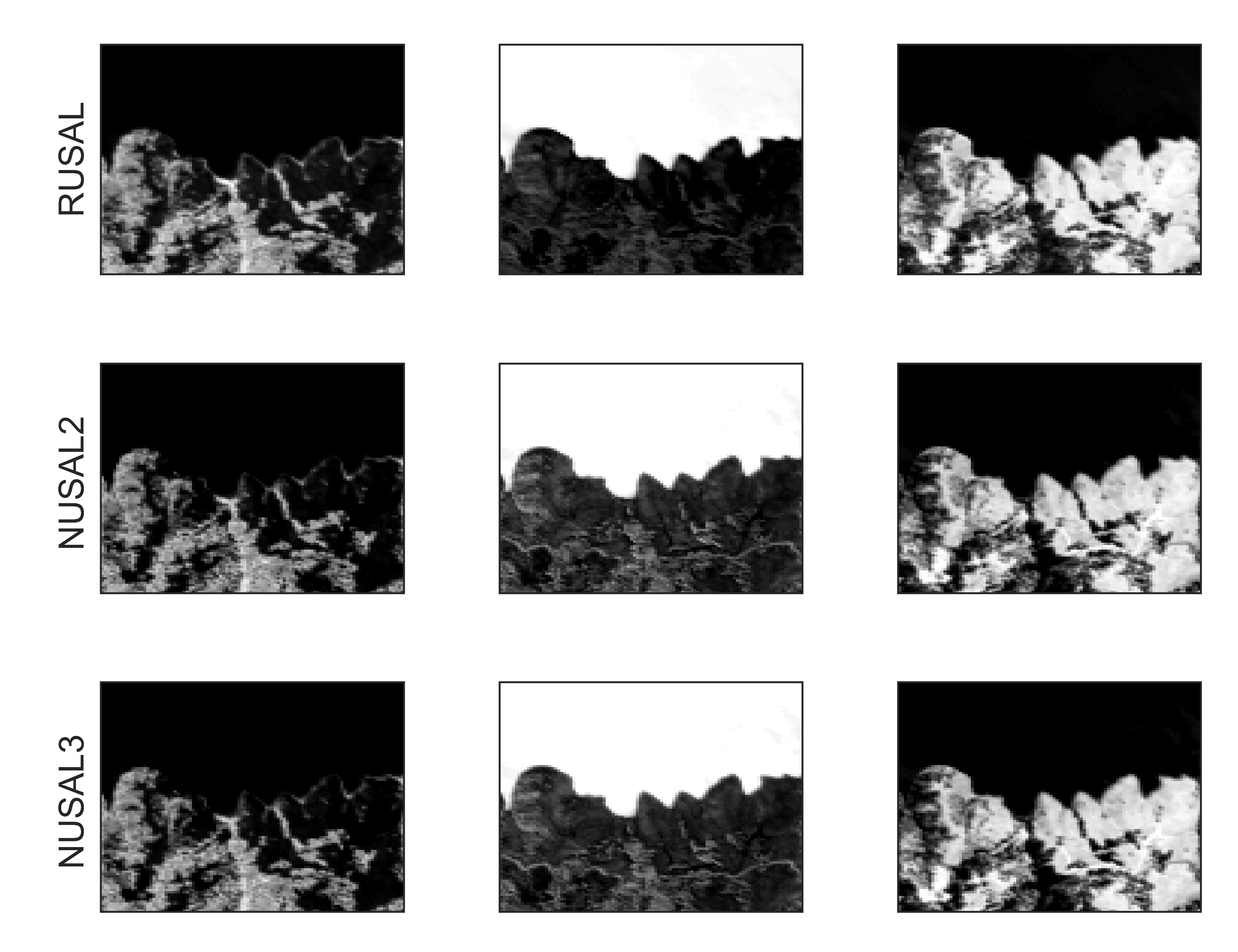}
\caption{Estimated abundance maps with different algorithms for the Moffett image.  (Left) vegetation, (middle) water, (right) soil.} \label{fig:Abundances_Moffett_Large_NUSAL_RUSAL}
\end{figure}

\begin{figure}[h!]
\centering
\includegraphics[width=0.95\figwidth,height=8cm]{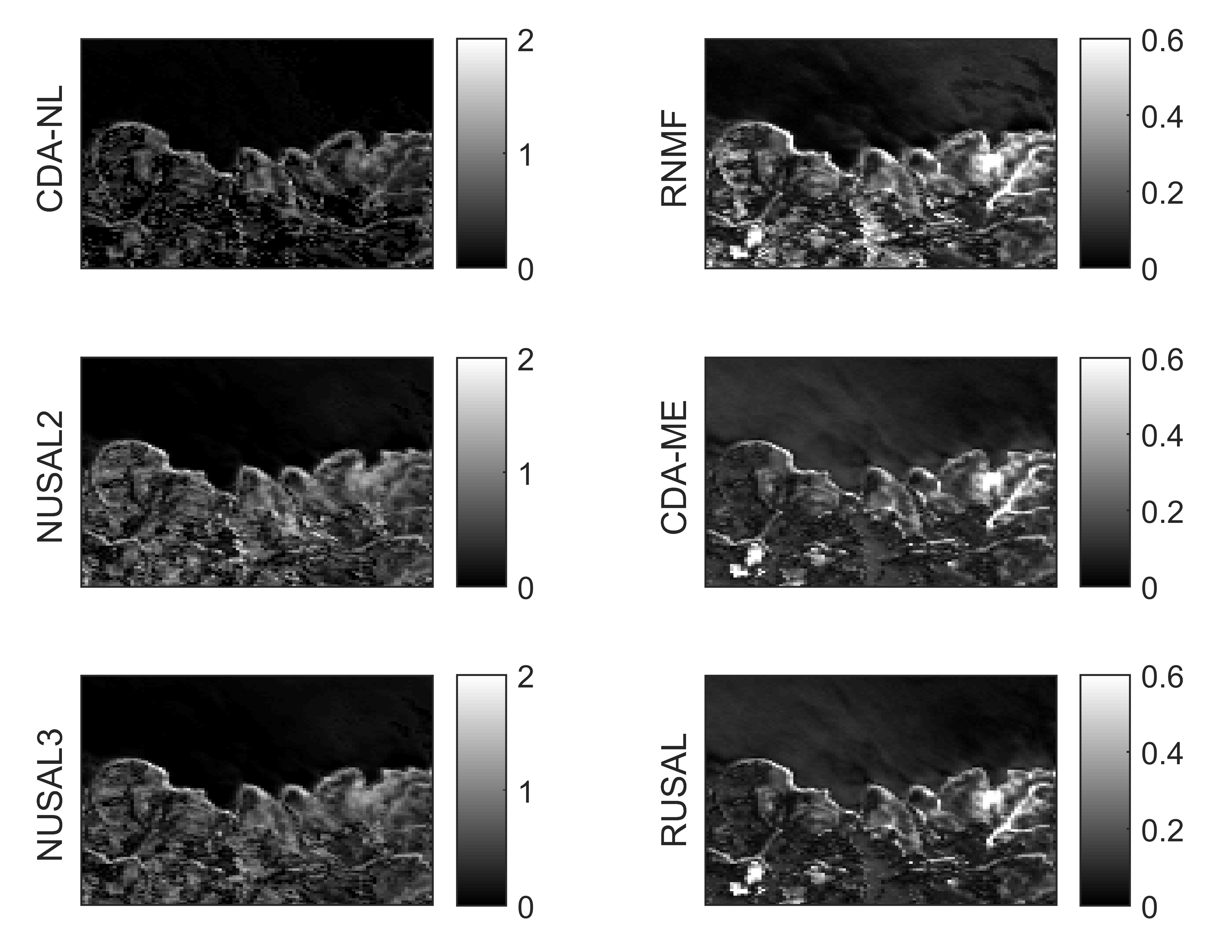}
\caption{Residual maps  for the Moffett image  obtained with $||\hat{\bsy}_{i,j} - \bsM \hat{\bsa}_{i,j}||$.} \label{fig:ResidMaps_Moffett_Large_NUSAL_RUSAL}
\end{figure}

\begin{figure}[h!]
\centering
\includegraphics[width=0.95\figwidth,height=8cm]{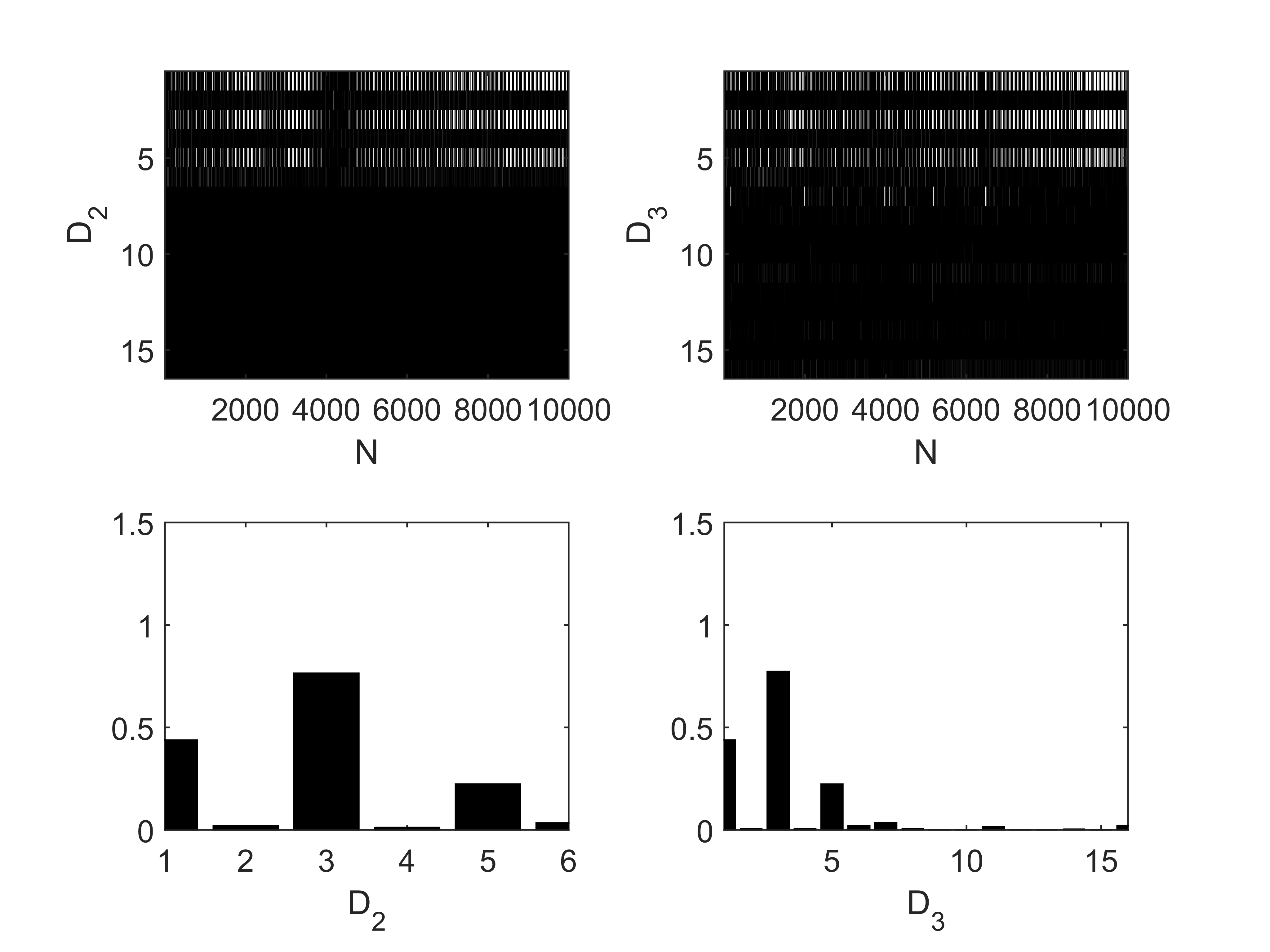}
\caption{Nonlinear coefficients obtained with NUSAL-$2$ and NUSAL-$3$ for the Moffett image. (Top) matrix ($D_k \times N$) of NL coefficients (the color scale is [0,1]), (bottom) averaged coefficient values of each nonlinear interaction term ($1/N \sum_{n=1}^{N}{\gamma_{n}^{(r,r')}},  \forall r,r'$).} \label{fig:NLCoeff_Moffett_Large_NUSAL_RUSAL}
\end{figure}

\begin{figure}[h!]
\centering
\includegraphics[width=0.95\figwidth,height=11cm]{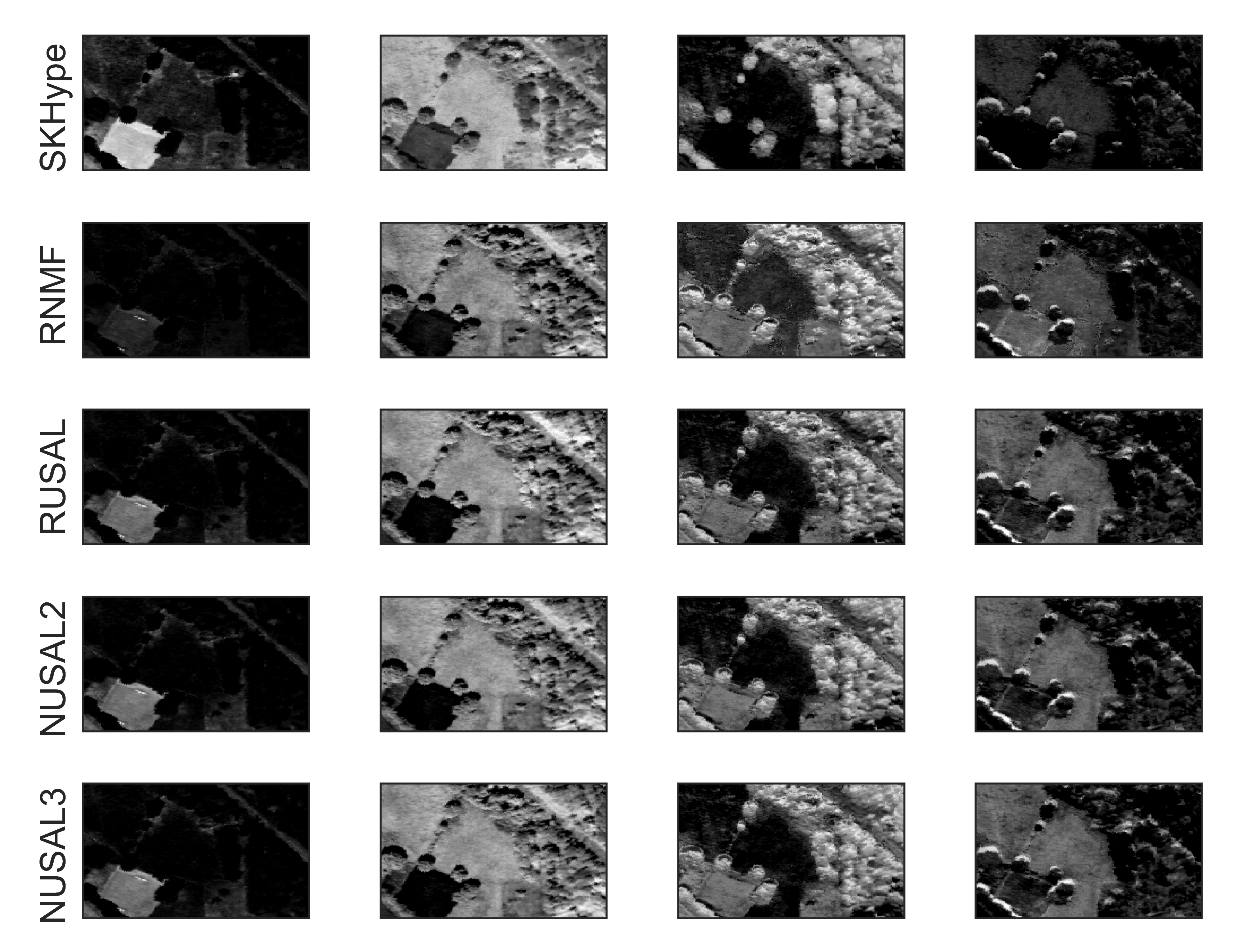}
\caption{Estimated abundance maps with different algorithms for the Madonna image.  from left to right: soil, grass, tree, shadow.} \label{fig:Abundances_MadonnaEV_Large_NUSAL_RUSAL}
\end{figure}

\begin{figure}[h!]
\centering
\includegraphics[width=0.95\figwidth,height=8cm]{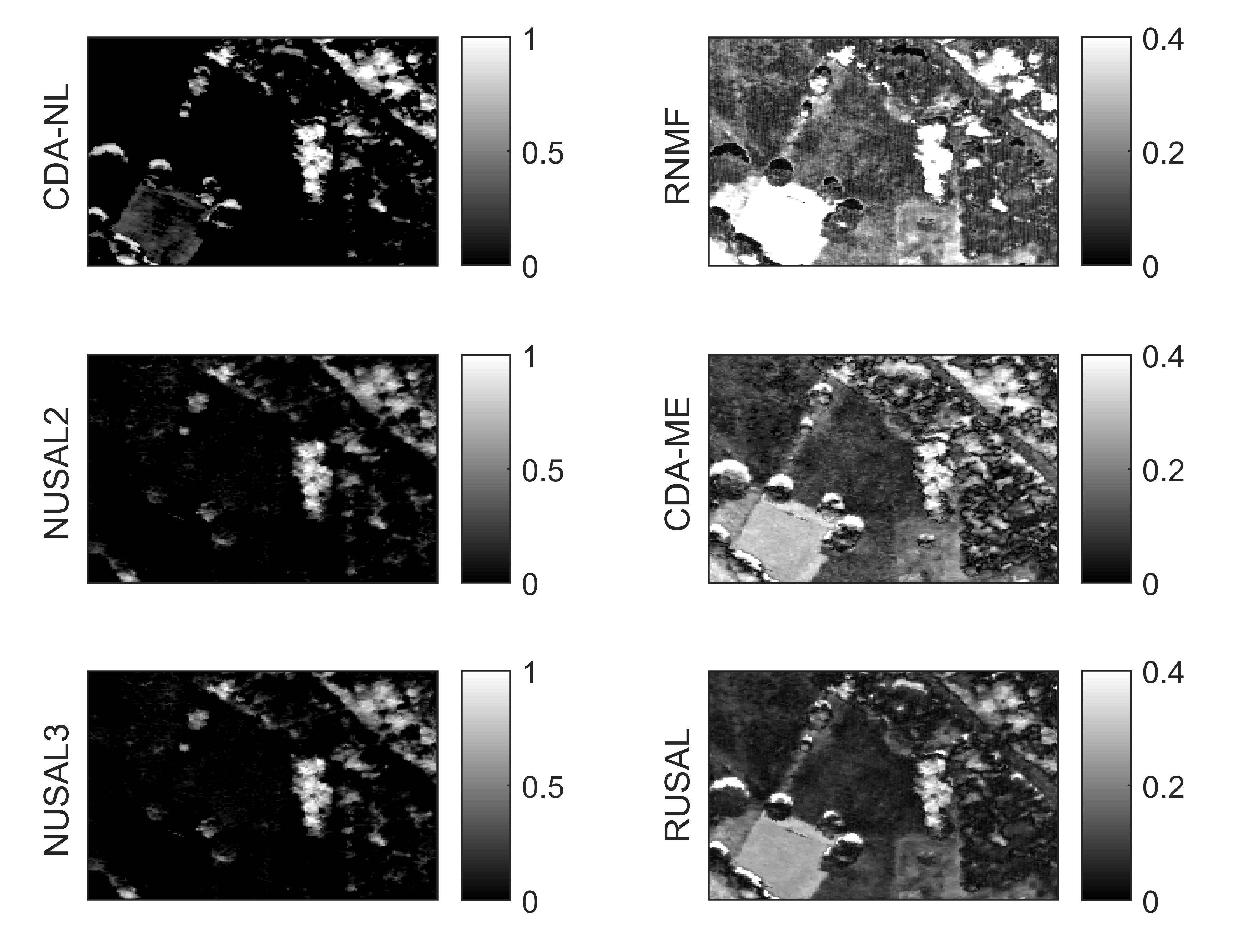}
\caption{Residual maps  for the Madonna image  obtained with $||\hat{\bsy}_{i,j} - \bsM \hat{\bsa}_{i,j}||$.} \label{fig:ResidMaps_MadonnaEV_Large_NUSAL_RUSAL}
\end{figure}

\begin{figure}[h!]
\centering
\includegraphics[width=0.95\figwidth,height=8cm]{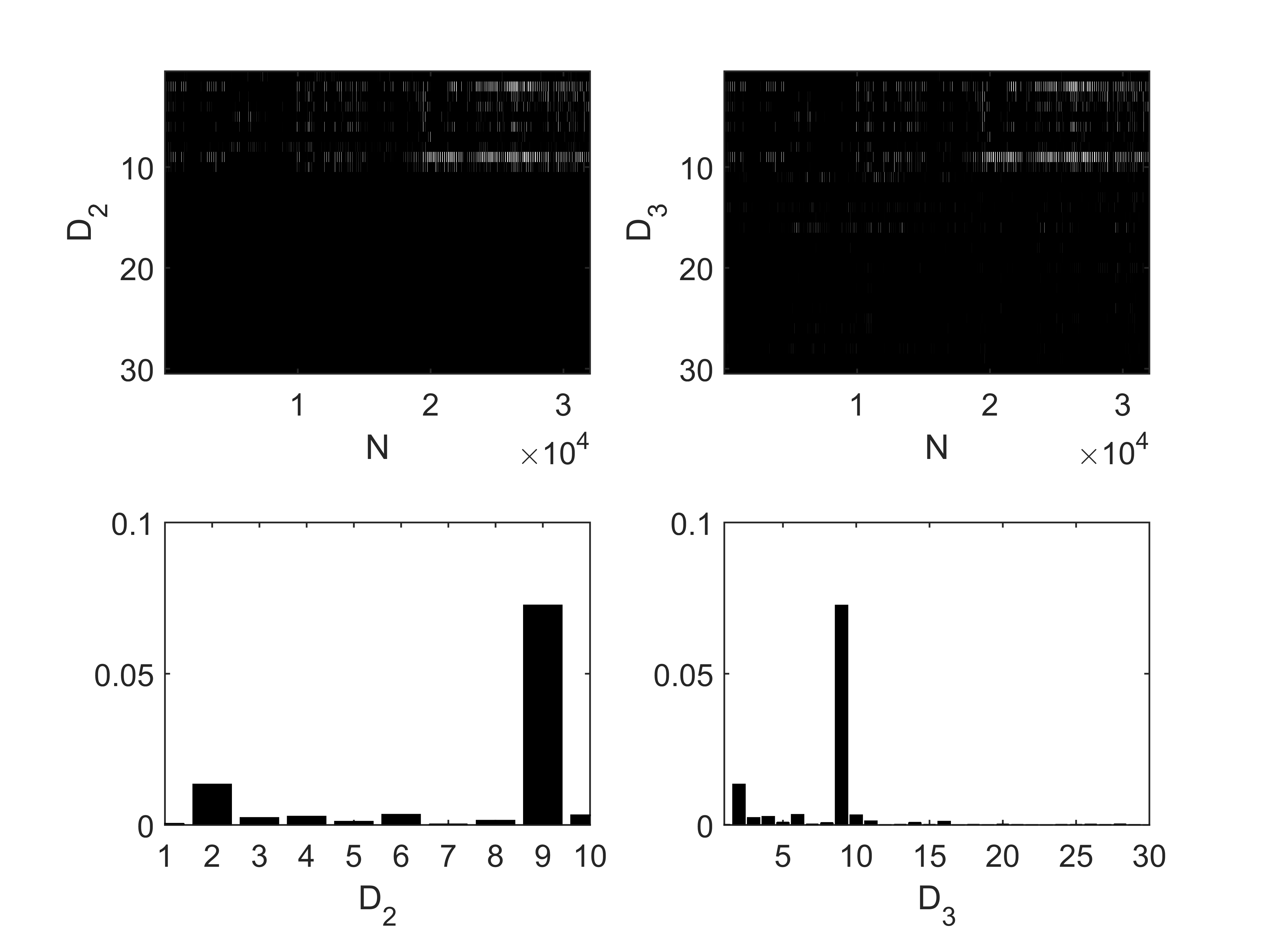}
\caption{Nonlinear coefficients obtained with NUSAL-$2$ and NUSAL-$3$ for the Madonna image.. (Top) matrix ($D_k \times N$) of NL coefficients (the color scale is [0,0.2]), (bottom) averaged coefficient values of each nonlinear interaction term ($1/N \sum_{n=1}^{N}{\gamma_{n}^{(r,r')}},  \forall r,r'$).} \label{fig:NLCoeff_MadonnaEV_Large_NUSAL_RUSAL}
\end{figure}

\begin{figure}[h!]
\centering
\includegraphics[width=0.95\figwidth,height=11cm]{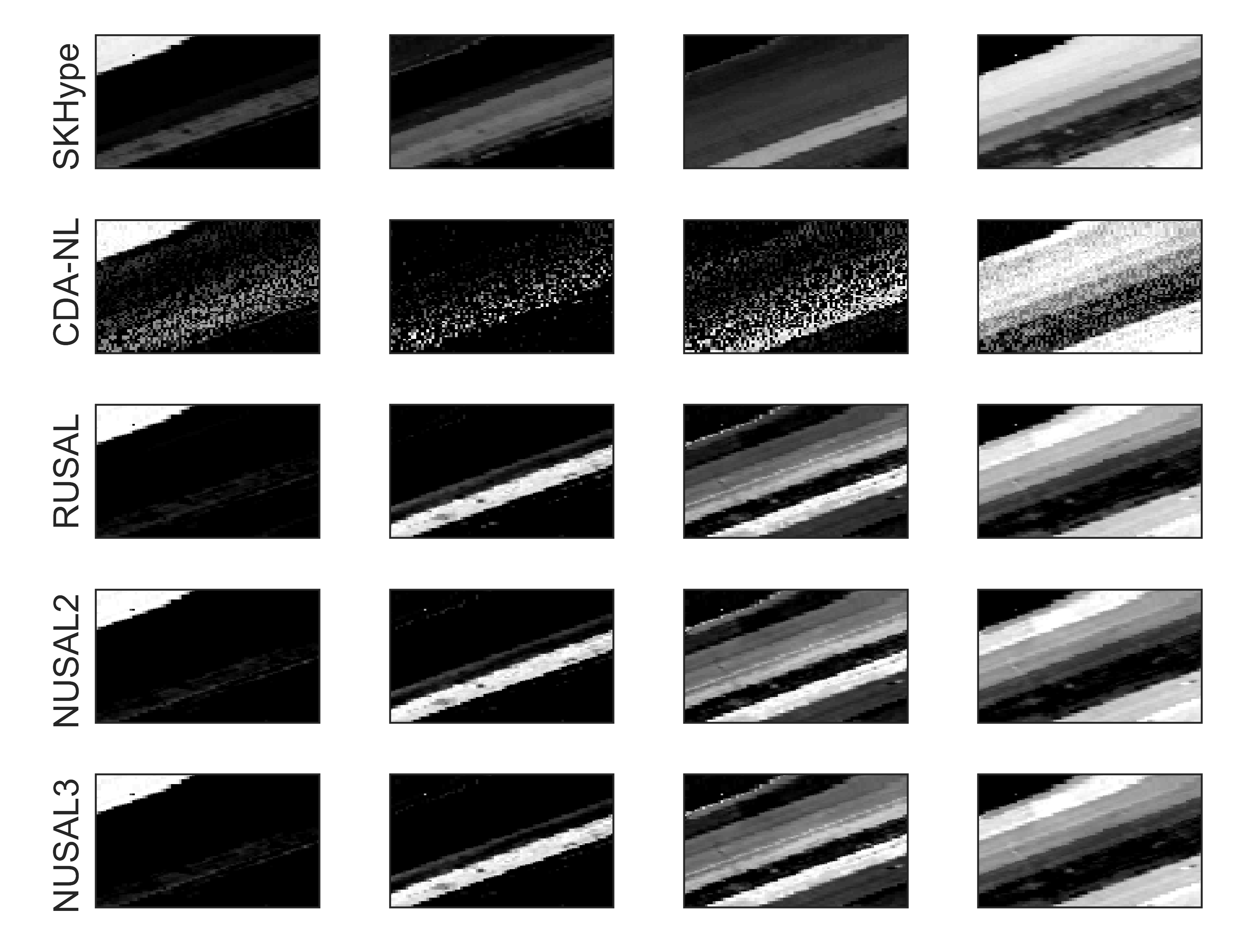}
\caption{Estimated abundance maps with different algorithms for the Salinas image. From left to right:  Corn$\underline{\;}$senesced$\underline{\;}$green$\underline{\;}$weeds + lettuce-4-5, Broccoli, lettuce-6, and lettuce-7.} \label{fig:Abundances_Salinas_NUSAL_RUSAL}
\end{figure}

\begin{figure}[h!]
\centering
\includegraphics[width=0.95\figwidth,height=8cm]{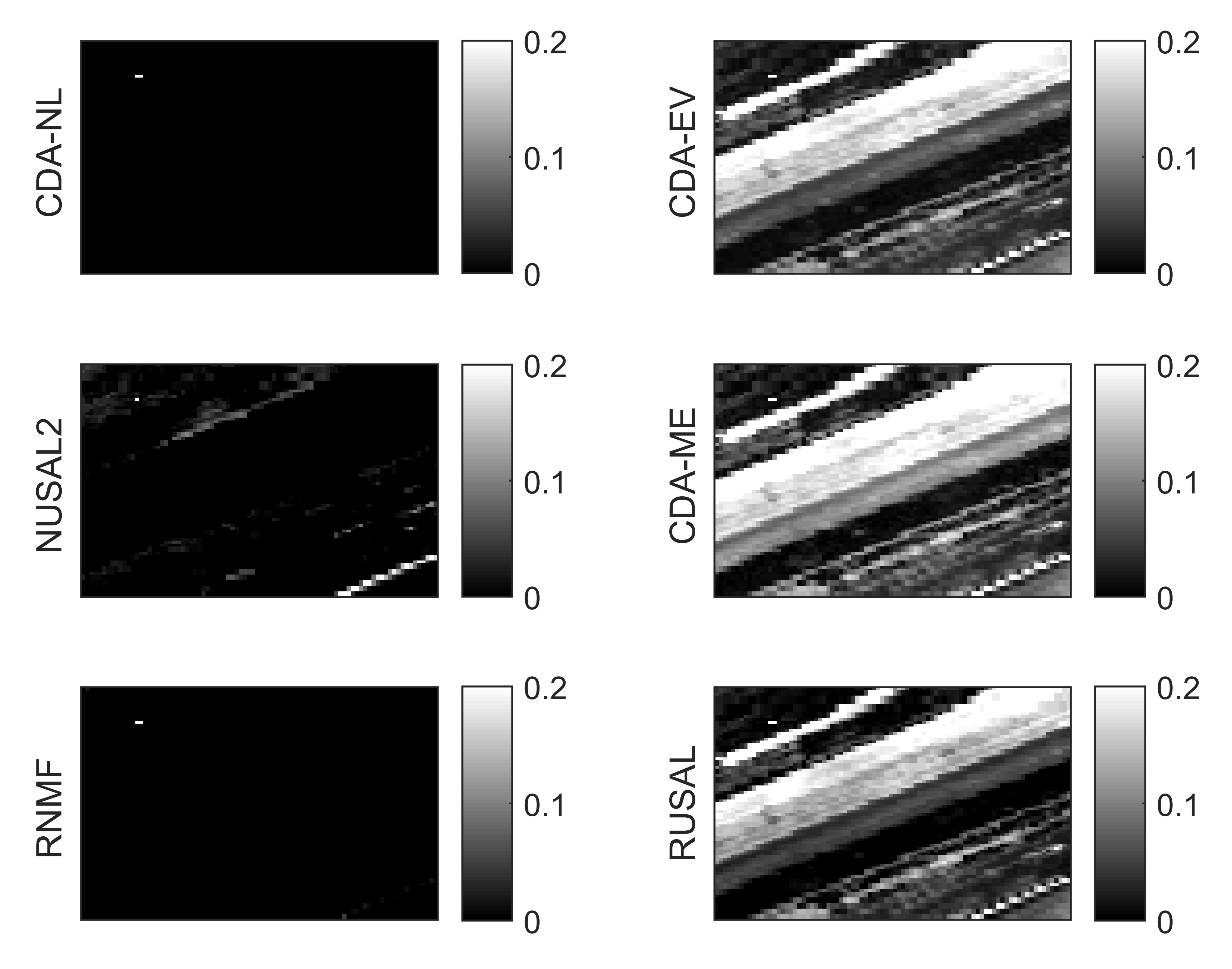}
\caption{Residual maps  for the Salinas image  obtained with $||\hat{\bsy}_{i,j} - \bsM \hat{\bsa}_{i,j}||$.} \label{fig:ResidMaps_Salinas_NUSAL_RUSAL}
\end{figure}

\begin{figure}[h!]
\centering
\includegraphics[width=1.0\figwidth,height=8cm]{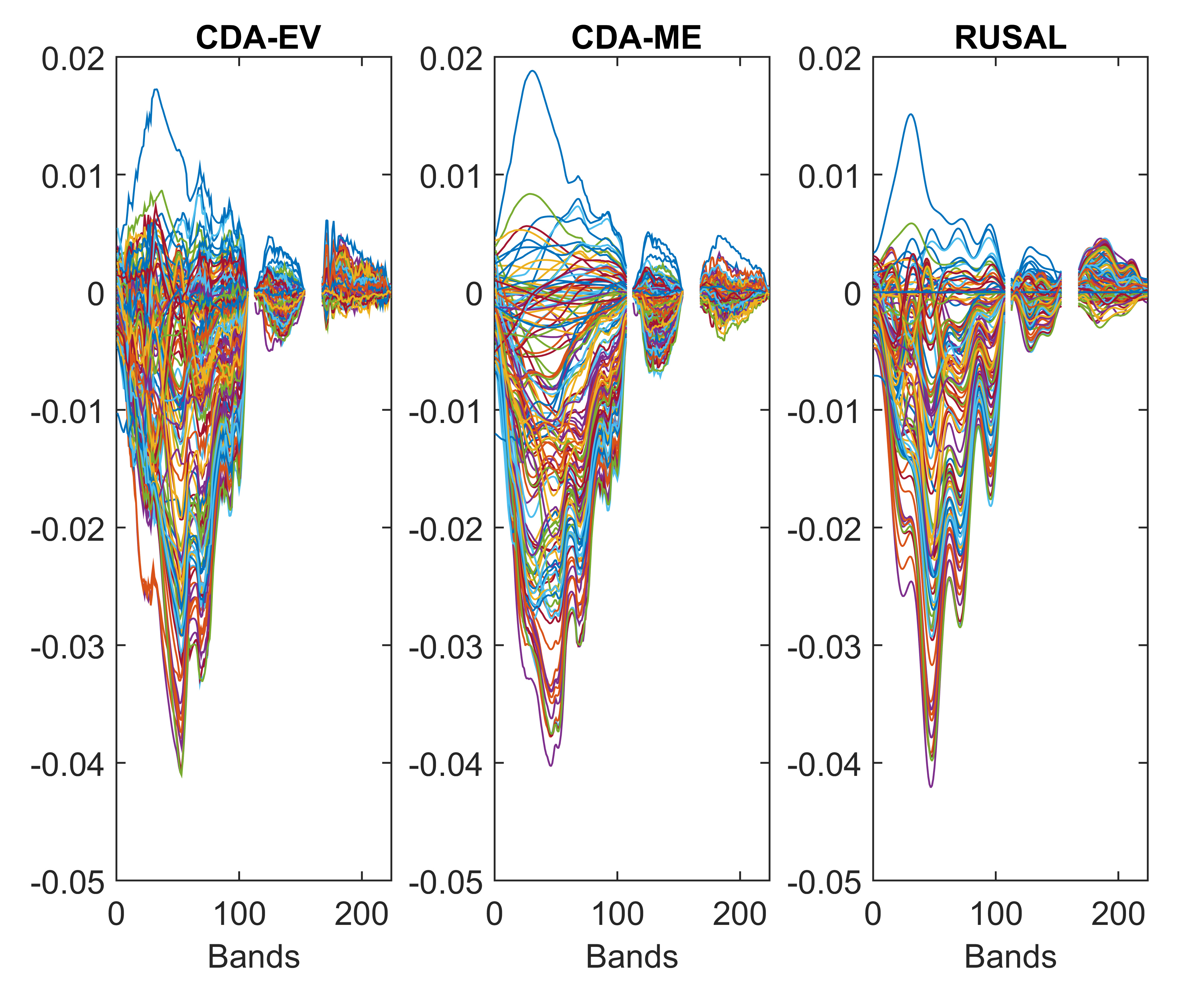}
\caption{Example of outlier spectra for the Salinas image obtained with (left) CDA-EV, (middle) CDA-ME and (right) RUSAL.} \label{fig:SpectraOut_Salinas_NUSAL_RUSAL}
\end{figure}

Table \ref{tab:Results_ImReal} shows the unmixing performance for the different algorithms. Overall, the NL and robust algorithms provide a better fit than the LMM-based ones. Among the sophisticated algorithms, the proposed NUSAL-$2$, NUSAL-$3$ and RUSAL algorithms provide the best performance for the computational cost. The algorithms all generated similar abundance maps for the Moffett image and we only show those of NUSAL-$2$, NUSAL-$3$ and RUSAL in Fig. \ref{fig:Abundances_Moffett_Large_NUSAL_RUSAL}, for brevity. Fig. \ref{fig:ResidMaps_Moffett_Large_NUSAL_RUSAL} presents the residual maps associated with the NL algorithms (left column) and the robust algorithms (right column). This figure highlights good agreement between the NL algorithms that detect nonlinearity  in the coastal region (as in \cite{Halimi2011TGRS}). In addition to this region, the robust algorithms (RNMF, ME, RUSAL) detect other mismodelling effects probably due to EV as already reported in \cite{HalimiTIP2016}. The NL coefficients estimated by NUSAL-$2$ and NUSAL-$3$ are reported in Fig. \ref{fig:NLCoeff_Moffett_Large_NUSAL_RUSAL}. This figure shows good agreement between the estimated bilinear coefficients when considering NUSAL-$2$ and NUSAL-$3$. Moreover, it  highlights the sparse behavior of the nonlinear coefficients and clearly shows that they are mainly due to the second order interactions. Indeed, Fig. \ref{fig:NLCoeff_Moffett_Large_NUSAL_RUSAL} (bottom-right) shows that the average of the nonlinear coefficients over all the pixels is higher for the first six terms, i.e., the second order terms. 
The  abundances  obtained  for  the  Madonna  scene  are  displayed  in Fig. \ref{fig:Abundances_MadonnaEV_Large_NUSAL_RUSAL} for SKhype, RNMF, RUSAL, NUSAL-$2$, and NUSAL-$3$ (the other algorithms provided similar maps to NUSAL/RUSAL and were not displayed for brevity).  This figure shows a slight difference between the RNMF soil map and the other algorithms. Similar differences are observed when considering the residual maps in Fig. \ref{fig:ResidMaps_MadonnaEV_Large_NUSAL_RUSAL} since RNMF detected a higher residual effect in the soil area (bottom-left corner in the RNMF image) than ME and RUSAL. Apart this, the robust algorithms detected residuals in the shadow areas and in trees. The latter is mainly due to the presence of multiple scattering effects as highlighted by the NL algorithms that show similar maps (see left figures). In a similar manner to Moffett field, the NUSAL-$2$ and NUSAL-$3$ estimated NL coefficients are mainly due to bilinear interactions as highlighted in Fig. \ref{fig:NLCoeff_MadonnaEV_Large_NUSAL_RUSAL}. This justifies the good behavior of the bilinear models  \cite{Altmann2012,Halimi2011TGRS,HalimiIGARSS2011,Nascimento2009,Fan2009,MeganemTGRS2014,Altmann2014} that assume that the effect of the interaction terms decreases when increasing the interaction orders. 
The abundances obtained for the Salinas scene are displayed in Fig. \ref{fig:Abundances_Salinas_NUSAL_RUSAL} for SKhype, CDA-NL, RUSAL, NUSAL-$2$, and NUSAL-$3$ (the other algorithms provided similar maps to NUSAL/RUSAL and were not displayed for brevity). Because of the high EV effect, both CDA-NL and SKhype fails to extract the abundances of Broccoli, and lettuce-6. The residual maps shown in Fig. \ref{fig:ResidMaps_Salinas_NUSAL_RUSAL} confirm this since NL algorithms detect a reduced effect while CDA-EV, CDA-ME and RUSAL detect more EV effect especially in the region of the lettuce. These results highlight the ability of CDA-ME and RUSAL to capture EV effects. Fig. \ref{fig:SpectraOut_Salinas_NUSAL_RUSAL} shows some randomly selected outlier spectra obtained with CDA-EV, CDA-ME and RUSAL algorithms. These spectra show a similar global shape while they highlight the properties of each algorithm. Indeed, it can be seen that the CDA-EV and CDA-ME algorithms provide rougher spectra that are more realistic than RUSAL.  However,  the RUSAL algorithm allows the absence of outliers (null spectra)  thanks to the sparsity promoting property imposed on the outliers.  
To summarize, the obtained results highlighted the benefit of RUSAL/NUSAL-$K$  that estimate abundance and residual maps
which are in good agreement with state-of-the-art algorithms, but at a lower computational cost. NUSAL-$K$ generalizes the common bilinear models and provide NL coefficient maps associated with different interaction orders. This provides a useful tool to better analyze the scattering effect between the physical elements. RUSAL provides a flexible tool to capture different mismodelling effects due to EV, NL or outliers. It is more robust than NUSAL-$K$ with respect to the variation of the endmember number $R$, but provide less information regarding the interaction terms. Table \ref{tab:Models} finally summarizes the main characteristics of the nonlinear and robust algorithms considered in this paper.   

\begin{table} \centering
\centering \caption{Results on real images. RE (resp. SAM) should be multiplied by $10^{-3}$ (resp. $\times 10^{-2}$). The time of processing the whole image is given in seconds.}
\begin{tabular}{|c|c|c||c|c||c|c|}
  \cline{2-7}
\multicolumn{1}{c|}{}   & \multicolumn{2}{c|}{Moffett } & \multicolumn{2}{|c|}{Madonna }  & \multicolumn{2}{|c|}{Salinas } \\
\cline{2-7}
\multicolumn{1}{c|}{} &  SAM & Time  &  SAM & Time &  SAM & Time \\
\cline{2-7}
\hline    FCLS     &  $12.7$ &   \blue{$1$} & $4.8$ &   \blue{$5$} & $3.8$ &   \blue{$1$}   \\
\hline    SUNSAL   &  $12.7$ &   \red{$0.1$} & $4.8$ &   \red{$0.5$} & $3.9$ &   \red{$0.1$}   \\
\hline    SKhype   &  - &   $177$ & - &   $551$ & - &   $136$         \\
\hline    CDA-NL   &  $10.7$ &   $317$ & $4.6$ &   $2700$ & $6.3$ &   $206$   \\
\hline    CDA-EV   &  $5.5$ &   $252$ & $3.3$ &   $2972$ & $1.8$ &   $27$         \\
\hline    CDA-ME   &  \red{$3.3$} &   $17$ & $3.1$ &   $342$ & \blue{$1.7$} &   $40$    \\
\hline    RNMF     &  $8.3$ &   $278$ & \red{$1.9$} &   $703$ & $3.2$ &   $97$      \\
\hline    RUSAL    &  \blue{$3.7$} &   $46$ & \blue{$2.6$} &   $94$ & \red{$1.6$} &   $20$   \\
\hline    NUSAL-$2$   &  $11.0$ &   $13$ & $4.6$ &   $94$ & $3.8$ &   $5$        \\
\hline    NUSAL-$3$   &  $10.4$ &   $29$ & $4.6$ &   $180$ & $3.8$ &   $19$\\
  \hline
\end{tabular}
\label{tab:Results_ImReal}
\end{table}

\setlength{\tabcolsep}{5pt}
\renewcommand{\arraystretch}{1.2}
\begin{table}[h] \centering
\centering \caption{Characteristics of the studied models/algorithms. ``Pos.'' stands for positivity,  ``Spat.'' for spatial, ``Spec.'' for spectral, ``Illumin.'' for Illumination, ``SM'' for smooth, ``SP'' for sparse, (+++) best results, and (+) good results.  }
\begin{tabular}{|c|c|c|c|c|c|c|c|}
\cline{2-8} \multicolumn{1}{c|}{}  &   Effects, & \multicolumn{3}{c|}{Residuals}  &  Illumin. &  Robust  & \multirow{2}{*}{Time} \\ 
\cline{3-5} \multicolumn{1}{c|}{}  &   LMM+     & Pos.     & Spat.     & Spec.    &  coeff. c &  to R       & \\ 
\hline  SKhype    &     NL               & \text{\sffamily X}   &  -        & -        & \text{\sffamily X}  & $\checkmark$ &  + \\ 
\hline  CDANL     &     NL-$2$                  & $\checkmark$ & SM & -           & $\checkmark$        & \text{\sffamily X} &  + \\ 
\hline  CDAEV     &     EV                      & \text{\sffamily X} & SM & SM    & \text{\sffamily X}  & \text{\sffamily X} &  + \\ 
\hline  CDAME     &     NL+EV                   & \text{\sffamily X} & SM & SM    & $\checkmark$        & $\checkmark$ &  ++ \\
\hline  RNMF      &     NL                      & $\checkmark$ & SP & -           & \text{\sffamily X}  & $\checkmark$ &  + \\
\hline  RUSAL     &     NL+EV                   & \text{\sffamily X} & SP & SM    & \text{\sffamily X}  & $\checkmark$ &  +++ \\
\hline  NUSAL-$K$     &     NL-$K$                  & $\checkmark$ & SP & -           & \text{\sffamily X}  & \text{\sffamily X} &  +++ \\	 
\hline 
\end{tabular}
\label{tab:Models}
\end{table}

\vspace{0.3cm}
\section{Conclusions} \label{sec:Conclusions} 
This paper has introduced two mixture models and their supervised unmixing algorithms. The two models accounted for the presence of nonlinearity or mismodelling effects by considering a residual term in addition to the linear mixture of endmembers.  The residual term was expressed as a sparse linear combination of some signals, thus, the proposed models reduced to a linear combination with respect to the abundances and the residual coefficients.  The unknown parameters associated with these models were estimated using an optimization approach that included convex regularization terms.  More precisely, the non-negativity and sum-to-one constraints were imposed on the abundances and the residual terms were assumed to be spatially sparse by considering a collaborative sparse regression approach.  The resulting convex problem was solved using an alternating direction method of multipliers whose convergence was theoretically ensured.  The proposed algorithms showed good performance when processing synthetic data generated with the linear model or other more sophisticated models. Results on real data confirmed the good performance of the proposed algorithms and showed their ability to extract different features in the observed scenes, with a reduced computational cost.  These results confirmed that most vegetation nonlinearity can be captured by bilinear interactions and that endmember variability is mainly located in vegetation areas. Future work includes the introduction of spatial correlation on the abundances. Considering endmember variability jointly with nonlinearity is also an interesting issue which would deserve to be investigated.


\appendices
\appendix[Derivations] \label{app:Complementary_derivations}

\subsection{Construction of $\bsQ^{(K)}$}\label{app:Construction_of_Q} 
Model  \eqref{eqt:GRCA_NL} requires the definition of the $\left(L\times D_K\right)$ matrix $\bsQ^{(K)}$ gathering the interaction spectra of all the orders lower than $K$. This section describes this matrix by providing its size and the coefficient of each interaction term. Before providing the full description of this matrix, let consider an example of $R=3$ endmembers and the matrix $\bsQ^{(K=3)}  = \left[\bsQ_2^{(3)},\bsQ_3^{(3)}\right]$. The number of interaction spectra is given by $16$ (see Table \ref{tab:D_k_examples}) while the corresponding spectra are given by concatenating the two matrices \\
  $\bsQ_2^{(3)} = \left(\sqrt{2} \bsm_{12}, \sqrt{2} \bsm_{13}, \sqrt{2} \bsm_{23}, \bsm_{11},\bsm_{22},\bsm_{33}\right),$ and	\\ $\bsQ_3^{(3)} = \left(\sqrt{3} \bsm_{112}
	, \sqrt{3} \bsm_{113}
	, \sqrt{3} \bsm_{122}
	, \sqrt{3} \bsm_{322} 
	, \sqrt{3} \bsm_{133}
	, \sqrt{3} \bsm_{233},  \right.$  $\left.
	 \sqrt{6} \bsm_{123},\bsm_{111},\bsm_{222},\bsm_{333}\right),$
with $\bsm_{ijk}  = \bsm_{i} \odot \bsm_{j} \odot \bsm_{k}$. For a formal mathematical description, denote $\bsQ^{(K)}=\left[\bsQ_2^{(K)},\bsQ_3^{(K)},\cdots,\bsQ_K^{(K)}\right]$, where $\bsQ_i^{(K)}$ gathers the interaction spectra of the $i$th order. The size of $\bsQ^{(K)}$ is then obtained by summing the size of the interaction spectra $D_{K}(i)$ associated with  the $i$th order, as follows
\begin{equation}
 D_K = \sum_{i=2}^{K}{ D_{K}(i)} = \sum_{i=2}^{K}{ \frac{\left(R+i-1\right)!}{i! \left(R-1\right)!}} \label{eqt:Dk}
\end{equation}  
where $x!= 1 \cdots (x-2) (x-1) x,$ denotes the factorial of $x$. Table \ref{tab:D_k_examples} shows some examples of $D_K$ for different values of $R$ and $K$. It is clear that increasing the interaction term $K$ leads to a fast increase of the number of interaction terms included in $\bsQ^{(K)}$. However, it is interesting to note that the sparsity promoting norms ($\ell_{1}$ and $\ell_{21}$) are well adapted to deal with large  $D_K$.
\begin{table}[h] \centering
\centering \caption{Example of $D_k$ for different values of $R$ and $K$.}
\begin{tabular}{|c|c|c|c|c|}
\cline{2-5} \multicolumn{1}{c|}{}    & $K=2$        &  $K=3$         & $K=4$  & $K=5$ \\
\hline     $R=3$      &  $  6$ & $ 16$   & $ 31$   & $  52$   \\
\hline     $R=6$      &  $ 21$ & $ 77$   & $203$   & $ 455$   \\
\hline     $R=10$     &  $ 55$ & $275$   & $990$   & $2992$   \\  
\hline 
\end{tabular}
\label{tab:D_k_examples}
\end{table}    
Similarly to  \cite{HalimiTIP2016,AltmannTCI2015b,ChenTSP2013}, each interaction term in  $\bsQ_i^{(K)}$ is weighted by a coefficient that is obtained by comparison with a homogeneous polynomial kernel of the $i$th degree. Straightforward computations show that the $i$th order spectra gathered in  $\bsQ_i^{(K)}$ are given by
\begin{equation}
\sqrt{\frac{i!}{\prod_{r=1}^{R} k_r!}}  \prod_{1\leq r \leq R}{\bsm_r^{k_r}}, \textrm{subject to} \sum_{r=1}^{R} k_r = i.
\label{eqt:Coeffs}
\end{equation}  

\subsection{ADMM algorithm}\label{app:ADMM_algorithm}

The list shown below provides details regarding the considered ADMM algorithm for both NUSAL and RUSAL. More precisely, we provide the solutions for the linear system of equations shown in line 8 of Algo. \ref{alg:ADMM_variant_for} and the MPO optimization problems shown in line 12. The details of the MPOs can be found, for example, in \cite{Combettes2011Book}.

\begin{itemize}
\item Linear system of equations: 	
			 {\small$$ \bsZ^{(k+1)} \leftarrow \bsG^{-1}   \sum_{j=1}^{J}{\left(\bsH_{j}\right)^{\top} \xi_{j}^{(k)}},$$} 
			\noindent with  $\bsG = \diag{[3\bone_{(1,R)},4\bone_{(1,D_K)}]}$ for the NL model and      $\bsG = 3\mathds{I}_{(R+D)}$  for the ME model	 \vspace{0.05cm}		
\item MPO for $g_1\left(\bsU_{1}\right) =    \mathcal{L}_{\bsP} \left(\bsU_{1}\right)$:  
      {\footnotesize$$\bsU_{1}^{(k+1)}   \leftarrow   \left\lbrace [\bsM,\bsP]^{\top} [\bsM,\bsP] + \mu \mathds{I}_{D+R} \right\rbrace^{-1}  \left\lbrace [\bsM,\bsP]^{\top} \bsY + \mu \bsV_{1}^{(k)} \right\rbrace $$}
\item MPO for $g_2\left(\bsU_{2}\right) =   \tau_1 {||\bsU_{2}||_1}$:  
			{\footnotesize$$ \bsU_{2}^{(k+1)}  \leftarrow   \textrm{soft} \left(\bsV_{2}^{(k)},\frac{\tau_1}{\mu}\right)$$}
\item MPO for $g_3\left(\bsU_{3}\right) =   \tau_2 {||\bsU_{3}||_{2,1}}$: 
		{\footnotesize$$\bsu_{3,n}^{(k+1)}   \leftarrow     \textrm{vect-soft} \left(\bsv_{3,n}^{(k)},\frac{\tau_2}{\mu}\right), \forall n$$}
\item MPO for $g_4\left(\bsU_{4}\right) =   \textit{i}_{\mathds{R}_{+}}\left(\bsU_{4}\right)$:
			{\footnotesize$$ \bsU_{4}^{(k+1)}  \leftarrow  \textrm{max}\left\lbrace\bsV_{4}^{(k)},0\right\rbrace $$} 
\item MPO for $g_5\left(\bsU_{5}\right) =  \textit{i}_{\left\lbrace\bone^{\top}\right\rbrace}\left(\bone^{\top}\bsU_{5} \right)$:
			{\footnotesize$$\bsU_{5}^{(k+1)}  \leftarrow  \left(\mathds{I}_{R}-\frac{1}{R} \bone_{(R,R)}\right) \bsV_{5}^{(k)} + \frac{1}{R} \bone_{(R,N)}$$}  
\end{itemize} 
where soft$(.)$ denotes the soft threshold operator given by $\textrm{soft} \left(\bsV,\frac{\tau}{\mu}\right)= \textrm{sign}(\bsV) \odot \textrm{max}\left\lbrace |\bsV|-\frac{\tau}{\mu}  ,0\right\rbrace$,  sign$(.)$ denotes the element-wise application of the sign function, $|\bsV|$ denotes the matrix of absolute values of the elements of $\bsV$, max$(.)$ is the element-wise maximum operator, and vect-soft$(.)$ is the well known vect-soft-threshold operator given by $\textrm{vect-soft} \left(\bsv,\frac{\tau}{\mu}\right)= 
\bsv \left( \frac{\textrm{max}\left\lbrace ||\bsv||_2-\frac{\tau}{\mu},0\right\rbrace }{\textrm{max}\left\lbrace ||\bsv||_2-\frac{\tau}{\mu},0\right\rbrace+\frac{\tau}{\mu}  }\right)$. Note finally that $\bsP=\bsQ$ for NUSAL and  $\bsP=\bsF^{\top}$ for RUSAL. 
 
\renewcommand{\baselinestretch}{1.4} 
\bibliographystyle{ieeetran}
\bibliography{biblio_all}

\begin{thebibliography}{10}
\providecommand{\url}[1]{#1}
\def\UrlFont{\rmfamily}
\providecommand{\newblock}{\relax}
\providecommand{\bibinfo}[2]{#2}
\providecommand\BIBentrySTDinterwordspacing{\spaceskip=0pt\relax}
\providecommand\BIBentryALTinterwordstretchfactor{4}
\providecommand\BIBentryALTinterwordspacing{\spaceskip=\fontdimen2\font plus
\BIBentryALTinterwordstretchfactor\fontdimen3\font minus
  \fontdimen4\font\relax}
\providecommand\BIBforeignlanguage[2]{{%
\expandafter\ifx\csname l@#1\endcsname\relax
\typeout{** WARNING: IEEEtran.bst: No hyphenation pattern has been}%
\typeout{** loaded for the language `#1'. Using the pattern for}%
\typeout{** the default language instead.}%
\else
\language=\csname l@#1\endcsname
\fi
#2}}

\bibitem{Bioucas2012}
J.~Bioucas-Dias, A.~Plaza, N.~Dobigeon, M.~Parente, Q.~Du, P.~Gader, and
  J.~Chanussot, ``Hyperspectral unmixing overview: Geometrical, statistical,
  and sparse regression-based approaches,'' \emph{IEEE J. Sel. Topics Appl.
  Earth Observat. Remote Sens.}, vol.~5, no.~2, pp. 354--379, April 2012.

\bibitem{HalimiTGRS2016}
A.~Halimi, P.~Honeine, M.~Kharouf, C.~Richard, and J.~Y. Tourneret,
  ``Estimating the intrinsic dimension of hyperspectral images using a
  noise-whitened eigengap approach,'' \emph{IEEE Trans. Geosci. Remote Sens.},
  vol.~54, no.~7, pp. 3811--3821, 2016.

\bibitem{Bioucas2008}
J.~M. {Bioucas-Dias} and J.~M.~P. Nascimento, ``Hyperspectral subspace
  identification,'' \emph{IEEE Trans. Geosci. Remote Sens.}, vol.~46, no.~8,
  pp. 2435--2445, Aug. 2008.

\bibitem{ChangTGRS2004}
C.~Chang and Q.~Du, ``Estimation of number of spectrally distinct signal
  sources in hyperspectral imagery,'' \emph{IEEE Trans. Geosci. Remote Sens.},
  vol.~42, no.~3, pp. 608--619, March 2004.

\bibitem{Nascimento2005}
J.~M.~P. Nascimento and J.~M. {Bioucas-Dias}, ``Vertex component analysis: A
  fast algorithm to unmix hyperspectral data,'' \emph{IEEE Trans. Geosci.
  Remote Sens.}, vol.~43, no.~4, pp. 898--910, April 2005.

\bibitem{Winter1999}
M.~Winter, ``Fast autonomous spectral end-member determination in hyperspectral
  data,'' in \emph{Proc. 13th Int. Conf. on Applied Geologic Remote Sensing},
  vol.~2, Vancouver, April 1999, pp. 337--344.

\bibitem{Heinz2001}
D.~C. Heinz and {C. -I Chang}, ``Fully constrained least-squares linear
  spectral mixture analysis method for material quantification in hyperspectral
  imagery,'' \emph{IEEE Trans. Geosci. Remote Sens.}, vol.~29, no.~3, pp.
  529--545, March 2001.

\bibitem{Chen2014}
J.~Chen, C.~Richard, and P.~Honeine, ``Nonlinear estimation of material
  abundances in hyperspectral images with $\ell _{1}$-norm spatial
  regularization,'' \emph{IEEE Trans. Geosci. Remote Sens.}, vol.~52, no.~5,
  pp. 2654--2665, May 2014.

\bibitem{BioucasWhispers2010}
J.~Bioucas-Dias and M.~Figueiredo, ``Alternating direction algorithms for
  constrained sparse regression: Application to hyperspectral unmixing,'' in
  \emph{Proc. IEEE GRSS Workshop on Hyperspectral Image and SIgnal Processing:
  Evolution in Remote Sensing (WHISPERS)}, June 2010, pp. 1--4.

\bibitem{AltmannTCI2015b}
Y.~Altmann, M.~Pereyra, and S.~McLaughlin, ``Bayesian nonlinear hyperspectral
  unmixing with spatial residual component analysis,'' \emph{IEEE Trans.
  Comput. Imaging}, vol.~1, no.~3, pp. 174--185, Sept 2015.

\bibitem{ChenTSP2013}
J.~Chen, C.~Richard, and P.~Honeine, ``Nonlinear unmixing of hyperspectral data
  based on a linear-mixture/nonlinear-fluctuation model,'' \emph{IEEE Trans.
  Signal Process.}, vol.~61, no.~2, pp. 480--492, Jan 2013.

\bibitem{Heylen2014}
R.~Heylen, M.~Parente, and P.~Gader, ``A review of nonlinear hyperspectral
  unmixing methods,'' \emph{IEEE J. Sel. Topics Appl. Earth Observat. Remote
  Sens.}, vol.~7, no.~6, pp. 1844--1868, June 2014.

\bibitem{Dobigeon2014}
N.~Dobigeon, J.-Y. Tourneret, C.~Richard, J.~Bermudez, S.~McLaughlin, and
  A.~Hero, ``Nonlinear unmixing of hyperspectral images: Models and
  algorithms,'' \emph{IEEE Signal Process. Mag.}, vol.~31, no.~1, pp. 82--94,
  Jan 2014.

\bibitem{Hapke1981}
B.~W. Hapke, ``Bidirectional reflectance spectroscopy. {I}. {T}heory,''
  \emph{J. Geophys. Res.}, vol.~86, pp. 3039--3054, 1981.

\bibitem{Altmann2012}
Y.~Altmann, A.~Halimi, N.~Dobigeon, and J.-Y. Tourneret, ``Supervised nonlinear
  spectral unmixing using a postnonlinear mixing model for hyperspectral
  imagery,'' \emph{IEEE Trans. Image Process.}, vol.~21, no.~6, pp. 3017--3025,
  June 2012.

\bibitem{Halimi2011TGRS}
A.~Halimi, Y.~Altmann, N.~Dobigeon, and J.-Y. Tourneret, ``Nonlinear unmixing
  of hyperspectral images using a generalized bilinear model,'' \emph{IEEE
  Trans. Geosci. Remote Sens.}, vol.~49, no.~11, pp. 4153--4162, 2011.

\bibitem{HalimiIGARSS2011}
------, ``Unmixing hyperspectral images using the generalized bilinear model,''
  in \emph{Proc. IEEE Int. Conf. Geosci. Remote Sens. (IGARSS)}, July 2011, pp.
  1886--1889.

\bibitem{Nascimento2009}
J.~M. {Bioucas-Dias} and J.~M.~P. Nascimento, ``Nonlinear mixture model for
  hyperspectral unmixing,'' in \emph{Proc. SPIE Image and Signal Processing for
  Remote Sensing XV}, L.~Bruzzone, C.~Notarnicola, and F.~Posa, Eds., vol.
  7477, no.~1.\hskip 1em plus 0.5em minus 0.4em\relax SPIE, 2009, p. 74770I.

\bibitem{Fan2009}
W.~Fan, B.~Hu, J.~Miller, and M.~Li, ``Comparative study between a new
  nonlinear model and common linear model for analysing laboratory
  simulated-forest hyperspectral data,'' \emph{International Journal of Remote
  Sensing}, vol.~30, no.~11, pp. 2951--2962, June 2009.

\bibitem{MeganemTGRS2014}
I.~Meganem, P.~Deliot, X.~Briottet, Y.~Deville, and S.~Hosseini,
  ``Linear-quadratic mixing model for reflectances in urban environments,''
  \emph{IEEE Trans. Geosci. Remote Sens.}, vol.~52, no.~1, pp. 544--558, Jan
  2014.

\bibitem{HalimiTIP2016}
A.~Halimi, P.~Honeine, and J.~M. Bioucas-Dias, ``Hyperspectral unmixing in
  presence of endmember variability, nonlinearity or mismodelling effects,''
  \emph{IEEE Trans. Image Process.}, 2016, to appear.

\bibitem{Somers2011}
B.~Somers, G.~P. Asner, L.~Tits, and P.~Coppin, ``Endmember variability in
  spectral mixture analysis: A review,'' \emph{Remote Sensing of Environment},
  vol. 115, no.~7, pp. 1603 -- 1616, 2011.

\bibitem{Zare2014}
A.~Zare and K.~Ho, ``Endmember variability in hyperspectral analysis:
  Addressing spectral variability during spectral unmixing,'' \emph{IEEE Signal
  Process. Mag.}, vol.~31, no.~1, pp. 95--104, Jan 2014.

\bibitem{AggarwalJSTARS2016}
H.~K. Aggarwal and A.~Majumdar, ``Hyperspectral unmixing in the presence of
  mixed noise using joint-sparsity and total variation,'' \emph{IEEE J. Sel.
  Topics Appl. Earth Observat. Remote Sens.}, 2016, to appear.

\bibitem{HeJSTARS2016}
W.~He, H.~Zhang, and L.~Zhang, ``Sparsity-regularized robust non-negative
  matrix factorization for hyperspectral unmixing,'' \emph{IEEE J. Sel. Topics
  Appl. Earth Observat. Remote Sens.}, pp. 1--13, 2016, to appear.

\bibitem{Eches2010ip}
O.~Eches, N.~Dobigeon, C.~Mailhes, and J.-Y. Tourneret, ``{B}ayesian estimation
  of linear mixtures using the normal compositional model. {A}pplication to
  hyperspectral imagery,'' \emph{IEEE Trans. Image Process.}, vol.~19, no.~6,
  pp. 1403--1413, June 2010.

\bibitem{Zare2013}
A.~Zare, P.~Gader, and G.~Casella, ``Sampling piecewise convex unmixing and
  endmember extraction,'' \emph{IEEE Trans. Geosci. Remote Sens.}, vol.~51,
  no.~3, pp. 1655--1665, March 2013.

\bibitem{Somers2012}
B.~Somers, M.~Zortea, A.~Plaza, and G.~Asner, ``Automated extraction of
  image-based endmember bundles for improved spectral unmixing,'' \emph{IEEE J.
  Sel. Topics Appl. Earth Observat. Remote Sens.}, vol.~5, no.~2, pp. 396--408,
  April 2012.

\bibitem{AltmannTCI2015}
Y.~Altmann, S.~McLaughlin, and A.~Hero, ``Robust linear spectral unmixing using
  anomaly detection,'' \emph{IEEE Trans. Comput. Imaging}, vol.~1, no.~2, pp.
  74--85, June 2015.

\bibitem{KalaitzisICML2012}
A.~A. Kalaitzis and N.~D. Lawrence, ``Residual components analysis,'' in
  \emph{Proc. ICML}, 2012, pp. 1--3.

\bibitem{Altmann2014}
Y.~Altmann, N.~Dobigeon, S.~McLaughlin, and J.-Y. Tourneret, ``Residual
  component analysis of hyperspectral images: Application to joint nonlinear
  unmixing and nonlinearity detection,'' \emph{IEEE Trans. Image Process.},
  vol.~23, no.~5, pp. 2148--2158, May 2014.

\bibitem{FevotteTIP2015}
C.~F\'evotte and N.~Dobigeon, ``Nonlinear hyperspectral unmixing with robust
  nonnegative matrix factorization,'' \emph{IEEE Trans. Image Process.},
  vol.~24, no.~12, pp. 4810--4819, June 2015.

\bibitem{SprechmannIEEETSP2011}
P.~Sprechmann, I.~Ramirez, G.~Sapiro, and Y.~C. Eldar, ``C-{H}ilasso: A
  collaborative hierarchical sparse modeling framework,'' \emph{IEEE Trans.
  Signal Process.}, vol.~59, no.~9, pp. 4183--4198, Sept 2011.

\bibitem{IordacheTGRS2014}
M.~D. Iordache, J.~M. Bioucas-Dias, and A.~Plaza, ``Collaborative sparse
  regression for hyperspectral unmixing,'' \emph{IEEE Trans. Geosci. Remote
  Sens.}, vol.~52, no.~1, pp. 341--354, Jan 2014.

\bibitem{IordacheTGRS2014b}
M.~D. Iordache, J.~M. Bioucas-Dias, A.~Plaza, and B.~Somers, ``{MUSIC-CSR}:
  Hyperspectral unmixing via multiple signal classification and collaborative
  sparse regression,'' \emph{IEEE Trans. Geosci. Remote Sens.}, vol.~52, no.~7,
  pp. 4364--4382, July 2014.

\bibitem{Afonso_TIP2011}
M.~Afonso, J.~Bioucas-Dias, and M.~Figueiredo, ``An augmented {L}agrangian
  approach to the constrained optimization formulation of imaging inverse
  problems,'' \emph{IEEE Trans. Image Process.}, vol.~20, no.~3, pp. 681--695,
  March 2011.

\bibitem{Boyd_FTML2011}
S.~Boyd, N.~Parikh, E.~Chu, B.~Peleato, and J.~Eckstein, ``Distributed
  optimization and statistical learning via the alternating direction method of
  multipliers,'' \emph{Found. Trends Mach. Learn.}, vol.~3, no.~1, pp. 1--122,
  Jan 2011.

\bibitem{AltmannICASSP2011}
Y.~Altmann, A.~Halimi, N.~Dobigeon, and J.~Y. Tourneret, ``Supervised nonlinear
  spectral unmixing using a polynomial post nonlinear model for hyperspectral
  imagery,'' in \emph{Proc. IEEE Int. Conf. Acoust., Speech, and Signal
  Process. (ICASSP)}, May 2011, pp. 1009--1012.

\bibitem{HeylenTGRS2016}
R.~Heylen and P.~Scheunders, ``A multilinear mixing model for nonlinear
  spectral unmixing,'' \emph{IEEE Trans. Geosci. Remote Sens.}, vol.~54, no.~1,
  pp. 240--251, Jan 2016.

\bibitem{HalimiSSPD2016}
A.~Halimi, Y.~Altmann, G.~S. Buller, S.~McLaughlin, W.~Oxford, D.~Clarke, and
  J.~Piper, ``Robust unmixing algorithms for hyperspectral imagery,'' in
  \emph{Sensor Signal Process. for Defence}, Sept. 2016, to appear.

\bibitem{Zhu_Arxiv2016}
F.~Zhu, A.~Halimi, P.~Honeine, B.~Chen, and N.~Zheng, ``Correntropy
  maximization via {ADMM} - application to robust hyperspectral unmixing,'' in
  \emph{ArXiv e-prints}, Feb. 2016.

\bibitem{Figueiredo_TIP2010}
M.~Figueiredo and J.~Bioucas-Dias, ``Restoration of poissonian images using
  alternating direction optimization,'' \emph{IEEE Trans. Image Process.},
  vol.~19, no.~12, pp. 3133--3145, Dec 2010.

\bibitem{EcksteinMP1992}
J.~Eckstein and D.~P. Bertsekas, ``On the {D}ouglas-{R}achford splitting method
  and the proximal point algorithm for maximal monotone operators,''
  \emph{Math. Programm.}, vol.~55, no.~1, pp. 293--318, 1992.

\bibitem{Combettes2011Book}
P.~L. Combettes and J.-C. Pesquet, \emph{Proximal Splitting Methods in Signal
  Processing}.\hskip 1em plus 0.5em minus 0.4em\relax New York, NY: Springer
  New York, 2011, pp. 185--212.

\bibitem{ENVImanual2003}
{{RSI} (Research Systems Inc.)}, \emph{ENVI User's guide Version 4.0}, Boulder,
  CO 80301 USA, Sept. 2003.

\bibitem{Dobigeon2008}
N.~Dobigeon, J.-Y. Tourneret, and {C.-I Chang}, ``Semi-supervised linear
  spectral unmixing using a hierarchical {B}ayesian model for hyperspectral
  imagery,'' \emph{IEEE Trans. Signal Process.}, vol.~56, no.~7, pp.
  2684--2695, July 2008.

\bibitem{Sheeren2011}
D.~Sheeren, M.~Fauvel, S.~Ladet, A.~Jacquin, G.~Bertoni, and A.~Gibon,
  ``Mapping ash tree colonization in an agricultural mountain landscape:
  {I}nvestigating the potential of hyperspectral imagery,'' in \emph{Proc. IEEE
  Int. Conf. Geosci. Remote Sens. (IGARSS)}, July 2011, pp. 3672--3675.

\bibitem{Halimi_TIP2015}
A.~Halimi, N.~Dobigeon, and J.-Y. Tourneret, ``Unsupervised unmixing of
  hyperspectral images accounting for endmember variability,'' \emph{IEEE
  Trans. Image Process.}, vol.~24, no.~12, pp. 4904--4917, 2015.

\bibitem{Altmann2014b}
Y.~Altmann, N.~Dobigeon, S.~McLaughlin, and J.-Y. Tourneret, ``Unsupervised
  post-nonlinear unmixing of hyperspectral images using a {H}amiltonian {M}onte
  {C}arlo algorithm,'' \emph{IEEE Trans. Image Process.}, vol.~23, no.~6, pp.
  2663--2675, June 2014.

\bibitem{LiTGRS2010}
J.~Li, J.~M. Bioucas-Dias, and A.~Plaza, ``Semisupervised hyperspectral image
  segmentation using multinomial logistic regression with active learning,''
  \emph{IEEE Trans. Geosci. Remote Sens.}, vol.~48, no.~11, pp. 4085--4098, Nov
  2010.

\bibitem{PlazaTGRS2005}
A.~Plaza, P.~Martinez, J.~Plaza, and R.~Perez, ``Dimensionality reduction and
  classification of hyperspectral image data using sequences of extended
  morphological transformations,'' \emph{IEEE Trans. Geosci. Remote Sens.},
  vol.~43, no.~3, pp. 466--479, March 2005.

\end{thebibliography}

\end{document}